# Predictions of SUSY Masses in the Minimal Supersymmetric Grand Unified Theory

W. de Boer[1] and R. Ehret[2]

*Inst. für Experimentelle Kernphysik, Univ. of Karlsruhe*
*Postfach 6980, D-76128 Karlsruhe 1, FRG*

and

D.I. Kazakov[3]

*Lab. of Theoretical Physics, Joint Inst. for Nuclear Research,*
*141 980 Dubna, Moscow Region, RUSSIA*

## Abstract

The Minimal Supersymmetric Standard Model (MSSM) distinguishes itself from other GUT's by a successful prediction of many unrelated phenomena with a minimum number of parameters.
Among them: a) Unification of the couplings constants; b) Unification of the masses; c) Proton decay; d) Electroweak symmetry breaking at a scale far below the unification scale and the corresponding relation between the gauge boson masses and the top quark mass.
A combined fit of the free parameters in the MSSM to these low energy constraints shows that the MSSM model can satisfy these constraints simultaneously. From the fitted parameters the masses of the as yet unobserved superpartners of the SM particles are predicted. The second order QCD coupling constant is required to be between 0.108 and 0.132. The complete second order renormalization group equations for the gauge and Yukawa couplings are used and analytical solutions for the neutral gauge boson, the Higgs masses and the sparticle masses are derived, taking into account the one-loop corrections to the Higgs potential.
It is shown that a top mass of $174 \pm 16$ GeV, as suggested recently by the CDF Collaboration, constrains the mixing angle between the Higgs doublets in the MSSM to: $1.2 < \tan\beta < 5.5$ at the 90% C.L.. The most probable value corresponds to $\tan\beta = 1.56$, which leads to a stop mass below the top mass. In this case the stop production in $p\bar{p}$ collisions would contribute to the top signature. This could be an explanation for the large effective $t\bar{t}$ cross section observed by CDF.

---
[1] Bitnet: DEBOERW@CERNVM
[2] Bitnet: BD21@DKAUNI2
[3] E-mail: KAZAKOVD@THSUN1.JINR.DUBNA.SU

# 1 Introduction

Grand Unified Theories (GUT) hold the promise of "explaining" the difference between the electromagnetic, weak and strong nuclear forces: their different strenghts are simply due to radiative corrections. Furthermore, they are candidates to explain several unrelated observations about our universe, e.g. they almost automatically lead to baryon number violation, thus providing a possible explanation for the matter-antimatter asymmetry in our universe [1] and the spontaneous symmetry breaking of the unified force into the known forces at a sufficient high energy can cause the inflationary scenario of the universe, thus providing an explanation for the origin of matter and the homogeneity of the universe on a large scale [2, 3].

The Grand Unification idea has been recently subjected to a new test using the new precise LEP data [4, 5, 6]. The result clearly indicates that the minimal Standard Model (SM), being extrapolated to 15 orders of magnitude does not lead to unification [5]. Indeed, the coupling constants, as measured precisely at LEP, do not become equal at a single energy if extrapolated to high energies from their values at $M_Z$. On the contrary, within the supersymmetric extension of the Standard Model, unification is achieved.

The Minimal Supersymmetric extension Standard Model (MSSM) has become the leading candidate for a Grand Unified Theory (GUT). Supersymmetry [7] presupposes a symmetry between fermions and bosons, thus introducing spin 0 partners of the quarks and leptons – called squarks and sleptons – and spin 1/2 partners of the gauge bosons and Higgs particles – called gauginos and Higgsinos. Since these predicted particles have not been observed sofar, these supersymmetric (SUSY) particles must be heavier than the known particles, implying that supersymmetry must be broken. However, from the unification condition a first estimate of the SUSY breaking scale could be made: it was found to be of the order of 1000 GeV, or more precisely $10^{3\pm1}$ GeV [5]. The uncertainty in this scale is mainly caused by the uncertainty in the strong coupling constant.

Assuming soft symmetry breaking at the GUT scale, all SUSY masses can be expressed in terms of 5 parameters and the masses at low energies are then determined by the well known Renormalization Group (RG) equations. So many parameters cannot be derived from the unification condition alone. Further constraints can be considered:

- $M_Z$ predicted from electroweak symmetry breaking [8] – [16].
- Constraints from the unification of Yukawa couplings [13, 14], [17] – [26].
- Constraints from the lower limit on the proton lifetime [27, 28, 29].
- Experimental lower limits on SUSY masses [30, 31].
- Constraints from the top mass suggested by CDF [32].

Of course, in many of the references given above, several constraints are studied simultaneously, since considering one constraint at a time yields only one relation between parameters. Trying to find complete solutions with only a few constraints requires then additional assumptions, like naturalness, no-scale models, fixed ratios for gaugino- and scalar masses or a fixed ratio for the Higgs mixing parameter and the scalar mass, or combinations of these assumptions.

Several ways to study the constraints simultaneously have been pursued. One can either sample the whole parameter space in a systematic or random way and check the regions which are allowed by the experimental constraints.

Alternatively, one can try a statistical analysis, in which all constraints are implemented in a $\chi^2$ definition and try to find the most probable region of the parameter space by minimizing the $\chi^2$ function.

In the first case one has to ask: which weight should one give to the various regions of parameter space and how large is the parameter space? Some sample the space logarithmically, thus emphasizing the low energy regions [15], others provide a linear sampling [23, 25, 33, 34, 35]. In the second case one is faced with the difficulty, that the function to be minimized is not monotonous, because of the experimental limits on the particle masses, proton lifetime and so on, since at the transitions where these constraints become effective, the derivative of the $\chi^2$ function is not defined. Fortunately, good minimizing programs in multidimensional parameter space, which do not rely on the derivatives, exist [36]. The advantage of such a statistical analysis is that one obtains probabilities for the allowed regions of the parameter space and can calculate confidence levels. The results of such an analysis will be presented after a short description of the experimental input values.

The requirement of electroweak symmetry breaking leads to a non-trivial minimum of the Higgs potential. Taking into account the one-loop radiative corrections proportional to the top Yukawa coupling we derive analytic expressions for the mass of the neutral gauge boson, the $Z^0$, as function of the mass of the top quark. These expressions can be used as strong constraints on the masses in the Higgs potential.

The paper is organized as follows: In sect. 2 we describe the MSSM together with the symmetry breaking part originated from supergravity and introduce 5 soft breaking parameters mentioned above. Sect. 3 contains the one-loop renormalization group equations for all the couplings and the parameters of the Higgs potential. One-loop analytic solutions to them are given. In sect. 4 we consider the leading radiative corrections to the Higgs potential and calculate their influence on the masses. Experimental constraints are discussed in sect. 5. The results are summarized in sect. 6.

For completeness we present in sections 2, 3 and 4 all the needed formulas used in our analysis, including those that are well known in the literature, since the notations usual are different and difficult to combine.

## 2 The Model

### 2.1 The Lagrangian

Minimal supersymmetric extention of the Standard Model is described by the Lagrangian containing the SUSY-symmetric part together with SUSY breaking terms originated from supergravity [7]. The breaking terms of the Lagrangian are given by:

$$\begin{aligned}
\mathcal{L}_{Breaking} &= -m_0^2 \sum_i |\varphi_i|^2 - m_{1/2} \sum_j \lambda_j \lambda_j \\
&\quad - Am_0 \left[ h_{ab}^u Q_a U_b^c H_2 + h_{ab}^d Q_a D_b^c H_1 + h_{ab}^e L_a E_b^c H_1 \right] - Bm_0 \left[ \mu H_1 H_2 \right],
\end{aligned} \quad (1)$$

Here

| | |
|---|---|
| $h_{ab}^{u,d,e}$ | are the Yukawa couplings, $a, b = 1, 2, 3$ run over the generations |
| $Q_a$ | are the SU(2) doublet squark fields |
| $U_a^c$ | are the SU(2) singlet charge-conjugated up-squark fields |
| $D_b^c$ | are the SU(2) singlet charge-conjugated down-squark fields |
| $L_a$ | are the SU(2) doublet slepton fields |
| $E_a^c$ | are the SU(2) singlet charge-conjugated slepton fields |
| $H_{1,2}$ | are the SU(2) doublet Higgs fields |
| $\varphi_i$ | are all scalar fields |
| $\lambda_j$ | are the gaugino fields, $j = 1, 2, 3$. |

From the Lagrangian given above it follows that at the GUT scale the squarks and sleptons have a common mass $m_0$ and the gauginos a common mass $m_{1/2}$. The full SUSY Lagrangian contains the following free parameters:

- 3 gauge couplings $\alpha_i$,
- the Yukawa couplings $h_i$,
- the Higgs fields mixing parameter $\mu$.

They are supplemented by the soft breaking ones:

- $m_0$, $m_{1/2}$, $A$, $B$, where A and B are the coupling constants for the Higgs fields.

These two sets of parameters determine completely the mass spectrum of all quarks, leptons, Higgs bosons and their superpartners.

## 2.2 Neutralino and Chargino Masses

The masses of the gauginos are denoted by $M_i$, where $i = 1, 2, 3$ corresponds to the U(1), SU(2) and SU(3) groups, respectively. For the gauginos of the SU(2) $\otimes$ U(1) group one has to consider the mixing with the Higgsinos, since they carry weak isospin, hypercharge and spin $\frac{1}{2}$ too. The mass terms in the full Lagrangian are [7]:

$$\mathcal{L}_{Gaugino-Higgsino} = -\frac{1}{2}M_3\bar{\lambda}_3^\alpha\lambda_3^\alpha - \frac{1}{2}\bar{\chi}M^{(0)}\chi - (\bar{\psi}M^{(c)}\psi + h.c.) \qquad (2)$$

where $\lambda_3^\alpha$ are the 8 Majorana gluino fields and

$$\chi = \begin{pmatrix} \tilde{B}^0 \\ \tilde{W}^3 \\ \tilde{H}_1^0 \\ \tilde{H}_2^0 \end{pmatrix}, \quad \psi = \begin{pmatrix} \tilde{W}^+ \\ \tilde{H}^+ \end{pmatrix},$$

are the Majorana neutralino and Dirac chargino fields. The mass matrices are:

$$M^{(0)} = \begin{pmatrix} M_1 & 0 & -M_Z \cos\beta \sin_W & M_Z \sin\beta \sin_W \\ 0 & M_2 & M_Z \cos\beta \cos_W & -M_Z \sin\beta \cos_W \\ -M_Z \cos\beta \sin_W & M_Z \cos\beta \cos_W & 0 & -\mu \\ M_Z \sin\beta \sin_W & -M_Z \sin\beta \cos_W & -\mu & 0 \end{pmatrix} \qquad (3)$$

$$M^{(c)} = \begin{pmatrix} M_2 & \sqrt{2}M_W \sin\beta \\ \sqrt{2}M_W \cos\beta & \mu \end{pmatrix} \qquad (4)$$

The last matrix has two chargino eigenstates $\tilde{\chi}^{\pm}_{1,2}$ with mass eigenvalues

$$M^2_{1,2} = \frac{1}{2}\left[M^2_2 + \mu^2 + 2M^2_W \right.$$
$$\left. \mp \sqrt{(M^2_2 - \mu^2)^2 + 4M^4_W \cos^2 2\beta + 4M^2_W(M^2_2 + \mu^2 + 2M_2\mu\sin 2\beta)}\right], \quad (5)$$

where at the GUT scale the masses of the gaugino fields of the U(1), SU(2) and SU(3) groups are equal to $m_{1/2}$, i. e. $M_1 = M_2 = M_3 = m_{1/2}$.

For the Higgs doublets

$$H_1(1,2,-\frac{1}{2}) = \begin{pmatrix} H^0_1 \\ H^-_1 \end{pmatrix}, \quad H_2(1,2,\frac{1}{2}) = \begin{pmatrix} H^+_2 \\ H^0_2 \end{pmatrix},$$

the tree level potential has the form:

$$V(H_1,H_2) = m^2_1|H_1|^2 + m^2_2|H_2|^2 - m^2_3(H_1H_2 + h.c.) + \frac{g^2 + g'^2}{8}(|H_1|^2 - |H_2|^2)^2. \quad (6)$$

The mass parameters in the potential fulfil the following boundary conditions at the GUT scale:

$$m^2_1 = m^2_2 = \mu^2_0 + m^2_0 \quad \text{and}$$
$$m^2_3 = -B\mu_0 m_0, \quad (7)$$

where $\mu_0$ is the value of $\mu$ at the GUT scale.

After spontaneous symmetry breaking the minimimum of the potential corresponds to non-zero vacuum expectation values of the Higgs fields:

$$<H_1> \equiv \begin{pmatrix} v_1 \\ 0 \end{pmatrix} = v\cos\beta, \quad <H_2> \equiv \begin{pmatrix} 0 \\ v_2 \end{pmatrix} = v\sin\beta, \quad (8)$$

$$\tan\beta \equiv \frac{v_2}{v_1}, \quad v^2 = v^2_1 + v^2_2 \quad (v \approx 174 \text{ GeV})$$

From the minimization of the potential one finds:

$$v^2 = \frac{4(m^2_1 - m^2_2\tan^2\beta)}{(g^2 + g'^2)(\tan^2\beta - 1)}, \quad \sin 2\beta = \frac{2m^2_3}{m^2_1 + m^2_2}, \quad (9)$$

which leads to the following expression for the masses:
Gauge bosons: $\qquad\qquad M^2_W = g^2 v^2/2, \quad M^2_Z = (g^2 + g'^2)v^2/2,$
CP-odd neutral Higgs $H_3 \equiv A$: $\qquad m^2_A = m^2_1 + m^2_2$
Charged Higgs masses $H^{\pm}$: $\qquad\qquad m^2_{H^{\pm}} = m^2_A + M^2_W$
CP-even neutral Higgs masses $H_{1,2}$:

$$m^2_{H_{1,2}} = \frac{1}{2}\left[m^2_A + M^2_Z \pm \sqrt{(m^2_A + M^2_Z)^2 - 4m^2_A M^2_Z \cos^2 2\beta}\right] \quad (10)$$

From eqs. (7) and (9) it follows that $\beta$ is related to $B$, and it is customary to use $\tan\beta$ instead of B as free parameter.

## 3 Renormalization Group Equations

In the previous section all masses were given at tree level. However, owing to radiative corrections, all couplings and masses become scale dependent (or running). Their scale dependence is described by the renormalization group (RG) equations, which depend on the particle content of the model in a given energy region. In the minimal subtraction scheme [37] one ignores the contribution of the particles heavier than the energy scale at hand. One can distinguish the following energy scales: $M_{\text{GUT}}, M_{\text{SUSY}}$, and $M_Z$, where $M_{\text{SUSY}}$ refers to some averaged mass of the superpartners.

For a not too large mixing angle $\tan\beta$ between the two Higgs doublets in the theory, as required by the present data (see section 5), the bottom Yukawa coupling is much smaller than the top Yukawa coupling. In this case the contribution from the bottom Yukawa coupling can be neglected, which allowed us to obtain analytical solutions for all superpartner masses.

### 3.1 The SUSY Mass Spectrum

In this section the first order RG equations and their solutions are summarized in order to get a complete mass spectrum as function of the parameters $\alpha_{\text{GUT}}$, $M_{\text{GUT}}$, $m_0$, $m_{1/2}$, $\mu$, $A$ and $\tan\beta$. The RG equations for the masses and the other SUSY breaking parameters in the region between $M_{\text{GUT}}$ and $M_{\text{SUSY}}$ are up to one-loop (we use the notation of the last paper of ref. [12]):

$$\begin{aligned}
\frac{d\tilde{m}_L^2}{dt} &= \left(3\tilde{\alpha}_2 M_2^2 + \frac{3}{5}\tilde{\alpha}_1 M_1^2\right) \\
\frac{d\tilde{m}_E^2}{dt} &= \left(\frac{12}{5}\tilde{\alpha}_1 M_1^2\right) \\
\frac{d\tilde{m}_Q^2}{dt} &= \left(\frac{16}{3}\tilde{\alpha}_3 M_3^2 + 3\tilde{\alpha}_2 M_2^2 + \frac{1}{15}\tilde{\alpha}_1 M_1^2\right) - \delta_{i3} Y_t(\tilde{m}_Q^2 + \tilde{m}_U^2 + m_2^2 + A_t^2 m_0^2 - \mu^2) \\
\frac{d\tilde{m}_U^2}{dt} &= \left(\frac{16}{3}\tilde{\alpha}_3 M_3^2 + \frac{16}{15}\tilde{\alpha}_1 M_1^2\right) - \delta_{i3} 2Y_t(\tilde{m}_Q^2 + \tilde{m}_U^2 + m_2^2 + A_t^2 m_0^2 - \mu^2) \\
\frac{d\tilde{m}_D^2}{dt} &= \left(\frac{16}{3}\tilde{\alpha}_3 M_3^2 + \frac{4}{15}\tilde{\alpha}_1 M_1^2\right) \qquad\qquad (11) \\
\frac{d\mu^2}{dt} &= 3(\tilde{\alpha}_2 + \frac{1}{5}\tilde{\alpha}_1 - Y_t)\mu^2 \\
\frac{dm_1^2}{dt} &= 3(\tilde{\alpha}_2 M_2^2 + \frac{1}{5}\tilde{\alpha}_1 M_1^2) + 3(\tilde{\alpha}_2 + \frac{1}{5}\tilde{\alpha}_1 - Y_t)\mu^2 \\
\frac{dm_2^2}{dt} &= 3(\tilde{\alpha}_2 M_2^2 + \frac{1}{5}\tilde{\alpha}_1 M_1^2) + 3(\tilde{\alpha}_2 + \frac{1}{5}\tilde{\alpha}_1)\mu^2 - 3Y_t(\tilde{m}_Q^2 + \tilde{m}_U^2 + m_2^2 + A_t^2 m_0^2) \\
\frac{dm_3^2}{dt} &= \frac{3}{2}(\tilde{\alpha}_2 + \frac{1}{5}\tilde{\alpha}_1 - Y_t)m_3^2 + 3\mu m_0 Y_t A_t - 3\mu(\tilde{\alpha}_2 M_2 + \frac{1}{5}\tilde{\alpha}_1 M_1) \\
\frac{dA_t}{dt} &= \left(\frac{16}{3}\tilde{\alpha}_3 \frac{M_3}{m_0} + 3\tilde{\alpha}_2 \frac{M_2}{m_0} + \frac{13}{15}\tilde{\alpha}_1 \frac{M_1}{m_0}\right) - 6Y_t A_t \\
\frac{dB}{dt} &= 3\left(\tilde{\alpha}_2 \frac{M_2}{m_0} + \frac{1}{5}\tilde{\alpha}_1 \frac{M_1}{m_0}\right) - 3Y_t A_t, \\
\frac{dM_i}{dt} &= -b_i \tilde{\alpha}_i M_i.
\end{aligned}$$

Here $m_U, m_D$ and $m_E$ refer to the masses of the superpartners of the quark and lepton singlets, while $m_Q$ and $m_L$ refer to the masses of the weak isospin doublet superpartners;

$m_1, m_2, m_3$ and $\mu$ are the mass parameters of the Higgs potential (see next section), while A and B are the couplings in $\mathcal{L}_{Breaking}$ as defined before; $M_i$ are the gaugino masses before any mixing.

The $\delta_{i3}$ term in the previous equations implies that only the Yukawa coupling of the third generation is taken into account. The top Yukawa coupling obeys

$$\frac{dY_t}{dt} = Y_t\left(\frac{16}{3}\tilde{\alpha}_3 + 3\tilde{\alpha}_2 + \frac{13}{15}\tilde{\alpha}_1\right) - 6Y_t^2, \tag{12}$$

while the RG equations for the gauge couplings in first order can be written as:

$$\frac{d\tilde{\alpha}_i}{dt} = -b_i\tilde{\alpha}_i^2. \tag{13}$$

The higher order corrections to these RG equations will be discussed in the next section.
Here

$$\tilde{\alpha}_i = \frac{\alpha_i}{4\pi}, \ Y_t = \frac{h_t^2}{(4\pi)^2}, \ t = \log(\frac{M_X^2}{Q^2}),$$

and the top Yukawa coupling $h_t$ is related to the top mass by

$$m_t = h_t(m_t)v\sin\beta.$$

Note that $m_t$ is the running mass. Its relation to the physical pole mass will be discussed in section 5.

The boundary conditions at $Q^2 = M_{\text{GUT}}^2$ or at $t = 0$ are:

$$\tilde{m}_Q^2 = \tilde{m}_U^2 = \tilde{m}_D^2 = \tilde{m}_L^2 = \tilde{m}_E^2 = m_0^2;$$

$$\mu^2 = \mu_0^2; \quad m_1^2 = m_2^2 = \mu_0^2 + m_0^2;$$

$$M_i = m_{1/2}; \quad \tilde{\alpha}_i(0) = \tilde{\alpha}_{\text{GUT}}, \quad i = 1, 2, 3$$

Solutions to the one-loop RG equations can be obtained analytically. In the MSSM they are [12]:

$$M_i(t) = \frac{\tilde{\alpha}_i(t)}{\tilde{\alpha}_i(0)}m_{1/2}$$

$$Y_t(t) = \frac{Y_0 E(t)}{1 + 6Y_0 F(t)} \tag{14}$$

$$A(t) = \frac{A_0}{1 + 6Y_0 F(t)} + \frac{m_{1/2}}{m_0}\left(H_2 - \frac{6Y_0 H_3}{1 + 6Y_0 F(t)}\right)$$

$$\mu^2(t) = \frac{\mu_0^2}{(1 + 6Y_0 F(t))^{1/2}}(1 + \beta_2 t)^{3/b_2}(1 + \beta_1 t)^{3/(5b_1)} \tag{15}$$

$$m_1^2(t) = m_0^2 + \mu^2(t) + m_{1/2}^2\tilde{\alpha}_{\text{GUT}}(\frac{3}{2}f_2(t) + \frac{3}{10}f_1(t))$$

$$m_2^2(t) = \mu^2(t) + m_{1/2}^2 e(t) + A_0 m_0 m_{1/2} f(t) + m_0^2(h(t) - k(t)A_0^2)$$

$$m_3^2(t) = q(t)m_3^2(0) + r(t)\mu_0 m_{1/2} + s(t)A_0 m_0 \mu_0$$

with the notation defined in the appendix.

For the light squark and slepton masses one finds [12]:

$$\begin{aligned}
\tilde{m}_{E_L}^2 &= m_0^2 + m_{1/2}^2 \tilde{\alpha}_{\text{GUT}} \left(\frac{3}{2}f_2(t) + \frac{3}{10}f_1(t)\right) - \cos(2\beta)M_Z^2(\frac{1}{2} - \sin^2\theta_W) \\
\tilde{m}_{\nu_L}^2 &= m_0^2 + m_{1/2}^2 \tilde{\alpha}_{\text{GUT}} \left(\frac{3}{2}f_2(t) + \frac{3}{10}f_1(t)\right) + \cos(2\beta)\frac{1}{2}M_Z^2 \\
\tilde{m}_{E_R}^2 &= m_0^2 + m_{1/2}^2 \tilde{\alpha}_{\text{GUT}} \left(\frac{6}{5}f_1(t)\right) - \cos(2\beta)M_Z^2 \sin^2\theta_W \\
\tilde{m}_{U_L}^2 &= m_0^2 + m_{1/2}^2 \tilde{\alpha}_{\text{GUT}} \left(\frac{8}{3}f_3(t) + \frac{3}{2}f_2(t) + \frac{1}{30}f_1(t)\right) - \cos(2\beta)M_Z^2(-\frac{1}{2} + \frac{2}{3}\sin^2\theta_W) \\
\tilde{m}_{D_L}^2 &= m_0^2 + m_{1/2}^2 \tilde{\alpha}_{\text{GUT}} \left(\frac{8}{3}f_3(t) + \frac{3}{2}f_2(t) + \frac{1}{30}f_1(t)\right) - \cos(2\beta)M_Z^2(\frac{1}{2} - \frac{1}{3}\sin^2\theta_W) \\
\tilde{m}_{U_R}^2 &= m_0^2 + m_{1/2}^2 \tilde{\alpha}_{\text{GUT}} \left(\frac{8}{3}f_3(t) + \frac{8}{15}f_1(t)\right) + \cos(2\beta)M_Z^2(\frac{2}{3}\sin^2\theta_W) \\
\tilde{m}_{D_R}^2 &= m_0^2 + m_{1/2}^2 \tilde{\alpha}_{\text{GUT}} \left(\frac{8}{3}f_3(t) + \frac{2}{15}f_1(t)\right) - \cos(2\beta)M_Z^2(\frac{1}{3}\sin^2\theta_W)
\end{aligned} \quad (16)$$

where $\beta$ is the mixing angle between the two Higgs doublets. For the third generation the effect of the top Yukawa coupling is taken into account through the $\delta_{i3}$ terms in eqns 11, in which case the solutions given above are changed. We found:

$$\begin{aligned}
\tilde{m}_{b_R}^2 &= \tilde{m}_{D_R}^2 \\
\tilde{m}_{b_L}^2 &= \tilde{m}_{D_L}^2 + \left[\frac{1}{3}(m_2^2 - \mu^2 - m_0^2) - \frac{1}{2}\tilde{\alpha}_{\text{GUT}}\left(f_2(t) + \frac{1}{5}f_1(t)\right)m_{1/2}^2\right] \\
\tilde{m}_{t_R}^2 &= \tilde{m}_{U_R}^2 + 2\left[\frac{1}{3}(m_2^2 - \mu^2 - m_0^2) - \frac{1}{2}\tilde{\alpha}_{\text{GUT}}\left(f_2(t) + \frac{1}{5}f_1(t)\right)m_{1/2}^2\right] + m_t^2 \\
\tilde{m}_{t_L}^2 &= \tilde{m}_{U_L}^2 + \left[\frac{1}{3}(m_2^2 - \mu^2 - m_0^2) - \frac{1}{2}\tilde{\alpha}_{\text{GUT}}\left(f_2(t) + \frac{1}{5}f_1(t)\right)m_{1/2}^2\right] + m_t^2
\end{aligned} \quad (17)$$

Note that only $b_L$ gets corrections from the top-quark Yukawa coupling through a loop with a chargino and a top-quark. The subscripts $L$ or $R$ do not indicate the helicity, since the squarks and sleptons have no spin. The labels just indicate in analogy to the non-SUSY particles, if they are $SU(2)$ doublets or singlets.

A non-negligible Yukawa coupling causes a mixing between the weak interaction eigenstates and the mass eigenstates. Since the mixing is proportional to the Yukawa coupling, we will only consider the mixing for the top quarks. The mass matrix is [12]:

$$\begin{pmatrix} \tilde{m}_{t_L}^2 & -m_t(A_t m_0 + \mu \cot\beta) \\ -m_t(A_t m_0 + \mu \cot\beta) & \tilde{m}_{t_R}^2 \end{pmatrix} \quad (18)$$

and the corresponding mass eigenstates are:

$$\tilde{m}_{t_{1,2}}^2 = \frac{1}{2}\left[\tilde{m}_{t_L}^2 + \tilde{m}_{t_R}^2 \mp \sqrt{(\tilde{m}_{t_L}^2 - \tilde{m}_{t_R}^2)^2 + 4m_t^2(A_t m_0 + \mu \cot\beta)^2}\right] \quad (19)$$

$\tilde{m}_{t_1}$ is defined as the lightest stop.

The functions $f_i(t)$ depend on $\tilde{\alpha}_{\text{GUT}}$ and $t = \ln(M_{\text{GUT}}^2/Q^2)$ as shown explicitly in the appendix. They were calculated for the parameters from the typical fit shown in table 5 ($\alpha_{\text{GUT}} = 1/24.3$, $M_{\text{GUT}} = 2.0 \cdot 10^{16}$ GeV, $\sin^2\theta_W = 0.2324$ and $A_t(0) = 0$). For these values one finds the following numerical formulae:

$$\begin{align}
\tilde{m}^2_{E_L}(t=66) &= m_0^2 + 0.52 m_{1/2}^2 - 0.27 \cos(2\beta) M_Z^2 \\
\tilde{m}^2_{\nu_L}(t=66) &= m_0^2 + 0.52 m_{1/2}^2 + 0.5 \cos(2\beta) M_Z^2 \\
\tilde{m}^2_{E_R}(t=66) &= m_0^2 + 0.15 m_{1/2}^2 - 0.23 \cos(2\beta) M_Z^2 \\
\tilde{m}^2_{U_L}(t=66) &= m_0^2 + 6.6 m_{1/2}^2 + 0.35 \cos(2\beta) M_Z^2 \\
\tilde{m}^2_{D_L}(t=66) &= m_0^2 + 6.6 m_{1/2}^2 - 0.42 \cos(2\beta) M_Z^2 \\
\tilde{m}^2_{U_R}(t=66) &= m_0^2 + 6.2 m_{1/2}^2 + 0.15 \cos(2\beta) M_Z^2 \\
\tilde{m}^2_{D_R}(t=66) &= m_0^2 + 6.1 m_{1/2}^2 - 0.07 \cos(2\beta) M_Z^2 \\
\tilde{m}^2_{b_R}(t=66) &= \tilde{m}^2_{D_R} \\
\tilde{m}^2_{b_L}(t=66) &= \tilde{m}^2_{D_L} - 0.48 m_0^2 - 1.21 m_{1/2}^2 \\
\tilde{m}^2_{t_R}(t=66) &= \tilde{m}^2_{U_R} + m_t^2 - 0.96 m_0^2 - 2.42 m_{1/2}^2 \\
\tilde{m}^2_{t_L}(t=66) &= \tilde{m}^2_{U_L} + m_t^2 - 0.48 m_0^2 - 1.21 m_{1/2}^2. \tag{20}
\end{align}$$

After mixing the mass eigenstates of the stop matrix are (using the same numerical input as for the light squarks):

$$\begin{align}
\tilde{m}^2_{t_{1,2}}(t=66) &= \frac{1}{2}\left[\tilde{m}^2_{t_L} + \tilde{m}^2_{t_R} \mp \sqrt{(\tilde{m}^2_{t_L} - \tilde{m}^2_{t_R})^2 + 4m_t^2(A_t m_0 + \mu/\tan\beta)^2}\right] \\
&\approx \frac{1}{2}\left[0.6 m_0^2 + 9.2 m_{1/2}^2 + 2m_t^2 - 0.19 \cos(2\beta) M_Z^2\right] \\
&\quad \mp \frac{1}{2}\sqrt{\left[1.6 m_{1/2}^2 + 0.5 m_0^2 - 0.5 \cos(2\beta) M_Z^2\right]^2 + 4 m_t^2 (A_t m_0 + \mu/\tan\beta)^2}. \tag{21}
\end{align}$$

where the values of $A_t$ and $\mu$ at the weak scale can be calculated as:

$$A_t(M_Z) = 4.6 A_t(0) + 1.7 \frac{m_{1/2}}{m_0} \tag{22}$$

$$\mu(M_Z) = 0.63 \mu_0 \tag{23}$$

Note that for large values of $A_t(0)$ or $\mu$ combined with a small $\tan\beta$ the splitting becomes large and one of the stop masses can become very small.

### 3.2 Second order RG Equations for the Couplings

After calculating the complete mass spectrum, one is able to check the unification conditions. Considering the wide range from $M_{\rm GUT}$ to $M_Z$, one needs the second order RG equations for the three couplings $\alpha_i$, the top Yukawa coupling $Y_t$ and the ratio $R_{b\tau} = m_b/m_\tau = \sqrt{Y_b/Y_\tau}$.

SUSY particles influence the evolution only through their appearance in the loops, so they enter only in higher order. Therefore it is sufficient to consider the loop corrections to the masses in first order, in which case simple analytical solutions can be found, as shown in the previous section. There is one exception: the corrections to the bottom and tau mass are compared directly with data, which implies that second order solutions have to be used for the RGE predicting the ratio of the bottom and tau mass. Since this ratio involves the top Yukawa coupling $Y_t$, the RGE for $Y_t$ has to be considered in second order too. These second order corrections are important for the bottom mass, since the strong coupling constant becomes large at the small scale of the bottom mass: $\alpha_s(m_b) \approx 0.2$.

| Particle | $b_1$ | $b_2$ | $b_3$ |
|---|---|---|---|
| $\tilde{g}$ | 0 | 0 | 2 |
| $\tilde{l}_l$ | $\frac{3}{10}$ | $\frac{1}{2}$ | 0 |
| $\tilde{l}_r$ | $\frac{3}{5}$ | 0 | 0 |
| $\tilde{w}$ | 0 | $\frac{4}{3}$ | 0 |
| $\tilde{q}-\tilde{t}$ | $\frac{49}{60}$ | 1 | $\frac{5}{3}$ |
| $\tilde{t}_l$ | $\frac{1}{60}$ | $\frac{1}{2}$ | $\frac{1}{6}$ |
| $\tilde{t}_r$ | $\frac{4}{15}$ | 0 | $\frac{1}{6}$ |
| $\tilde{h}$ | $\frac{2}{5}$ | $\frac{2}{3}$ | 0 |
| $H$ | $\frac{1}{10}$ | $\frac{1}{6}$ | 0 |
| $t$ | $\frac{17}{30}$ | 1 | $\frac{2}{3}$ |
| SM | $\frac{41}{10}$ | $-\frac{19}{6}$ | $-7$ |
| MSSM | $\frac{33}{5}$ | 1 | $-3$ |

| Particle | $b_{ij}$ |
|---|---|
| $\tilde{g}$ | $\begin{pmatrix} 0 & 0 & 0 \\ 0 & 0 & 0 \\ 0 & 0 & 48 \end{pmatrix}$ |
| $\tilde{w}$ | $\begin{pmatrix} 0 & 0 & 0 \\ 0 & \frac{64}{3} & 0 \\ 0 & 0 & 0 \end{pmatrix}$ |
| $\tilde{q}, \tilde{l},$ gauginos | $\begin{pmatrix} \frac{19}{15} & \frac{3}{5} & \frac{44}{15} \\ \frac{1}{5} & -\frac{7}{3} & 4 \\ \frac{11}{30} & \frac{3}{2} & -\frac{8}{3} \end{pmatrix}$ |
| Heavy Higgses and Higgsinos | $\begin{pmatrix} \frac{9}{50} & \frac{9}{10} & 0 \\ \frac{3}{10} & \frac{29}{6} & 0 \\ 0 & 0 & 0 \end{pmatrix}$ |

Table 1: Contributions to first order RGE coefficients.

Table 2: Contributions to second order RGE coefficients.

Defining a vector $\vec{\tilde{\alpha}} = \tilde{\alpha}_{i=1,..,4} = (\tilde{\alpha}_1, \tilde{\alpha}_2, \tilde{\alpha}_3, Y_t)$ the RG equations in second order can be written in a compact form:

$$\frac{d\tilde{\alpha}_i}{dt} = -b_i \tilde{\alpha}_i^2 - \tilde{\alpha}_i^2 \left( \sum_{j=1}^{3} b_{ij} \tilde{\alpha}_j - a_i Y_t \right), \quad i=1,2,3 \quad (24)$$

$$\frac{dY_t}{dt} = Y_t \sum_{i=1}^{4} \left( c_i \tilde{\alpha}_i - \sum_{j \geq i}^{4} c_{ij} \tilde{\alpha}_i \tilde{\alpha}_j \right) \quad (25)$$

$$\frac{dR_{b\tau}}{dt} = R_{b\tau} \sum_{i=1}^{4} \left( d_i \tilde{\alpha}_i - \sum_{j \geq i}^{4} d_{ij} \tilde{\alpha}_i \tilde{\alpha}_j \right) \quad (26)$$

Since there are no analytic solutions for the two-loop RG equations, we solve this system of 5 coupled differential equations numerically.

All coefficients depend on the particle content of the theory. Running the RG equations from $M_{\rm GUT}$ down to $M_Z$, the coefficients will be updated every time a particle or sparticle threshold is passed. So we start with the MSSM values and the contribution of $m_i$ to the coefficients is subtracted, as soon as the scale $Q$ is below $m_i$. The contributions of the various particles to $b_i$ and $b_{ij}$ have been summarized in tables 1 and 2 (from [19, 38, 39, 40, 41, 42]).

| Region I | $c_i$ | $c_{ij}$ | | | |
|---|---|---|---|---|---|
| | | $j=1$ | $j=2$ | $j=3$ | $j=4$ |
| $i=1$ | $\frac{13}{15}$ | $(\frac{13}{15}b_1 + \frac{169}{450})$ | 1 | $\frac{136}{45}$ | $\frac{6}{5}$ |
| $i=2$ | 3 | 0 | $(3b_2 + \frac{9}{2})$ | 8 | 6 |
| $i=3$ | $\frac{16}{3}$ | 0 | 0 | $(\frac{16}{3}b_3 + \frac{128}{9})$ | 16 |
| $i=4$ | $-6$ | 0 | 0 | 0 | $-22$ |
| Region II | | | | | |
| $i=1$ | $\frac{17}{20}$ | $\frac{1187}{600}$ | $-\frac{9}{20}$ | $\frac{19}{15}$ | $\frac{393}{80}$ |
| $i=2$ | $\frac{9}{4}$ | 0 | $-\frac{23}{4}$ | 9 | $\frac{225}{16}$ |
| $i=3$ | 8 | 0 | 0 | $-108$ | 36 |
| $i=4$ | $-\frac{9}{2}$ | 0 | 0 | 0 | $-12$ |

Table 3: One and two loop coefficients $c_i$ and $c_{ij}$ of the RG equation for the top Yukawa coupling $Y_t$, eq. (25). The region of integration is divided into two parts: $M_{\rm SUSY} < m < M_{\rm GUT}$, $m_t < m < M_{\rm SUSY}$. Below the top mass all coefficients $c_i$ and $c_{ij}$ are zero.

The total contributions from the SM particles can be summarized as:

$$a_i = \begin{pmatrix} a_1 \\ a_2 \\ a_3 \end{pmatrix} = \begin{pmatrix} 17/10 \\ 3/2 \\ 2 \end{pmatrix} \qquad (27)$$

$$b_i = \begin{pmatrix} b_1 \\ b_2 \\ b_3 \end{pmatrix} = \begin{pmatrix} 0 \\ -22/3 \\ -11 \end{pmatrix} + N_{Fam} \begin{pmatrix} 4/3 \\ 4/3 \\ 4/3 \end{pmatrix} + N_{Higgs} \begin{pmatrix} 1/10 \\ 1/6 \\ 0 \end{pmatrix}, \qquad (28)$$

while for the MSSM one finds:

$$a_i = \begin{pmatrix} a_1 \\ a_2 \\ a_3 \end{pmatrix} = \begin{pmatrix} 26/5 \\ 6 \\ 4 \end{pmatrix} \qquad (29)$$

$$b_i = \begin{pmatrix} b_1 \\ b_2 \\ b_3 \end{pmatrix} = \begin{pmatrix} 0 \\ -6 \\ -9 \end{pmatrix} + N_{Fam} \begin{pmatrix} 2 \\ 2 \\ 2 \end{pmatrix} + N_{Higgs} \begin{pmatrix} 3/10 \\ 1/2 \\ 0 \end{pmatrix}, \qquad (30)$$

Here $N_{Fam}$ is the number of families of matter supermultiplets and $N_{Higgs}$ is the number of Higgs doublets. We use $N_{Fam} = 3$ and $N_{Higgs} = 1$ or 2, which corresponds to the minimal SM or minimal SUSY model, respectively. Below the top mass $a_i = 0$.

| Region I | $d_i$ | $d_{ij}$ | | | |
|---|---|---|---|---|---|
| | | $j=1$ | $j=2$ | $j=3$ | $j=4$ |
| $i=1$ | $-\frac{2}{3}$ | $-(\frac{2}{3}b_1 + \frac{34}{45})$ | $-\frac{2}{5}$ | $\frac{4}{9}$ | $\frac{2}{5}$ |
| $i=2$ | $0$ | $0$ | $0$ | $4$ | $0$ |
| $i=3$ | $\frac{8}{3}$ | $0$ | $0$ | $(\frac{8}{3}b_3 + \frac{64}{9})$ | $0$ |
| $i=4$ | $-\frac{1}{2}$ | $0$ | $0$ | $0$ | $-\frac{5}{2}$ |
| Region II | | | | | |
| $i=1$ | $-1$ | $-\frac{53}{15}$ | $-\frac{27}{20}$ | $\frac{31}{30}$ | $-\frac{79}{160}$ |
| $i=2$ | $0$ | $0$ | $0$ | $\frac{9}{2}$ | $\frac{9}{32}$ |
| $i=3$ | $4$ | $0$ | $0$ | $-54$ | $-8$ |
| $i=4$ | $\frac{3}{4}$ | $0$ | $0$ | $0$ | $\frac{13}{4}$ |
| Region III | | | | | |
| $i=1$ | $-1$ | $-\frac{53}{15}$ | $-\frac{27}{20}$ | $\frac{31}{30}$ | $0$ |
| $i=2$ | $0$ | $0$ | $0$ | $\frac{9}{2}$ | $0$ |
| $i=3$ | $4$ | $0$ | $0$ | $-54$ | $0$ |
| $i=4$ | $0$ | $0$ | $0$ | $0$ | $0$ |
| Region IV | | | | | |
| $i=1$ | $-1$ | $-\frac{53}{15}$ | | $\frac{31}{30}$ | $0$ |
| $i=2$ | $0$ | | $0$ | | $0$ |
| $i=3$ | $4$ | | | $-54$ | $0$ |
| $i=4$ | $0$ | | | | $0$ |

Table 4: One and two loop coefficients $d_i$ and $d_{ij}$ of the RG equation for the ratio $R_{b\tau}$, eq. (26). The region of integration is divided into four parts: $M_{\rm SUSY} < m < M_{\rm GUT}$, $m_t < m < M_{\rm SUSY}$, $M_Z < m < m_t$, $m_b < m < m_Z$. The running of $m_\tau$ below $m_b$ is ignored. In these four regions the superpartners are decoupled in regions II, III and IV, the top in regions III and IV, and the Z and W in region IV. With the superpartners one has to be careful, since some are light and still survive in region II.

The $b_{ij}$'s for the standard $SU(3) \otimes SU(2) \otimes U(1)$ are:

$$b_{ij} = \begin{pmatrix} 0 & 0 & 0 \\ 0 & -\frac{136}{3} & 0 \\ 0 & 0 & -102 \end{pmatrix} + N_{Fam} \begin{pmatrix} \frac{19}{15} & \frac{3}{5} & \frac{44}{15} \\ \frac{1}{5} & \frac{49}{3} & 4 \\ \frac{11}{30} & \frac{3}{2} & \frac{76}{3} \end{pmatrix} + N_{Higgs} \begin{pmatrix} \frac{9}{50} & \frac{9}{10} & 0 \\ \frac{3}{10} & \frac{13}{6} & 0 \\ 0 & 0 & 0 \end{pmatrix}. \qquad (31)$$

For the SUSY model they become:

$$b_{ij} = \begin{pmatrix} 0 & 0 & 0 \\ 0 & -24 & 0 \\ 0 & 0 & -54 \end{pmatrix} + N_{Fam} \begin{pmatrix} \frac{38}{15} & \frac{6}{5} & \frac{88}{15} \\ \frac{2}{5} & 14 & 8 \\ \frac{11}{15} & 3 & \frac{68}{3} \end{pmatrix} + N_{Higgs} \begin{pmatrix} \frac{9}{50} & \frac{9}{10} & 0 \\ \frac{3}{10} & \frac{7}{2} & 0 \\ 0 & 0 & 0 \end{pmatrix}. \qquad (32)$$

Note that the $\alpha_i^2(\mu)$ coefficients are changed when taking the second order contributions into account and the running of each $\alpha_i$ depends on the values of the two other coupling constants.

The second order effects are small, because the $b_i$'s are multiplied by $\alpha_i/4\pi \leq 0.01$. Higher orders are presumably even smaller.

For the coefficients $c_i, c_{ij}$ of eq. (25) and $d_i, d_{ij}$ of eq. (26), we consider four regions; region (I) from $M_{\text{GUT}}$ to $M_{\text{SUSY}}$, region (II) from $M_{\text{SUSY}}$ to $m_{top}$, region (III) from $m_{top}$ to $M_Z$ and region (IV) from $M_Z$ to $m_b$; $M_{\text{SUSY}}$ is some averaged mass of superpartners. The coefficients in these different regions are shown in tables 3 and 4.

## 4 Radiative Corrections to the Higgs Potential

If one keeps only the radiative corrections proportional to the top Yukawa coupling, the one-loop effective Higgs potential can be written as [23, 43, 44, 45, 46, 47]:

$$
\begin{aligned}
V(H_1, H_2) &= m_1^2 |H_1|^2 + m_2^2 |H_2|^2 - m_3^2 (H_1 H_2 + h.c.) + \frac{g^2 + g^{'2}}{8} (|H_1|^2 - |H_2|^2)^2 \\
&+ \frac{3}{32\pi^2} \left[ \tilde{m}_{t_1}^4 (\ln \frac{\tilde{m}_{t_1}^2}{Q^2} - \frac{3}{2}) + \tilde{m}_{t_2}^4 (\ln \frac{\tilde{m}_{t_2}^2}{Q^2} - \frac{3}{2}) - 2 m_t^4 (\ln \frac{m_t^2}{Q^2} - \frac{3}{2}) \right],
\end{aligned} \quad (33)
$$

where $\tilde{m}_{t_i}$ and $m_t$ are field dependent masses:

$$m_t^2 = h_t^2 \, |H_2^0|^2 = h_t^2 v^2 \sin^2 \beta, \quad (34)$$

and $\tilde{m}_{t_1}^2$ and $\tilde{m}_{t_2}^2$ are the eigenvalues of the $2 \times 2$ $\tilde{t}_L$ and $\tilde{t}_R$ mass-squared matrix, given in eq. (19).

With $v_1, v_2$ defined in eq. 8 the one-loop minimization conditions are:

$$
\begin{aligned}
\frac{\partial V}{\partial H_1} &= 2 m_1^2 v_1 - 2 m_3^2 v_2 + \frac{g^2 + g^{'2}}{2} (v_1^2 - v_2^2) v_1 \\
&+ \frac{3}{8\pi^2} h_t^2 \mu (A_t m_0 v_2 + \mu v_1) \frac{b(\tilde{m}_{t1}^2) - b(\tilde{m}_{t2}^2)}{\tilde{m}_{t1}^2 - \tilde{m}_{t2}^2} = 0
\end{aligned} \quad (35)
$$

$$
\begin{aligned}
\frac{\partial V}{\partial H_2} &= 2 m_2^2 v_2 - 2 m_3^2 v_1 - \frac{g^2 + g^{'2}}{2} (v_1^2 - v_2^2) v_2 \\
&+ \frac{3}{8\pi^2} \left\{ h_t^2 A_t m_0 (A_t m_0 v_2 + \mu v_1) \frac{b(\tilde{m}_{t1}^2) - b(\tilde{m}_{t2}^2)}{\tilde{m}_{t1}^2 - \tilde{m}_{t2}^2} \right. \\
&+ \left. [(b(\tilde{m}_{t1}^2) + b(\tilde{m}_{t2}^2) - 2 b(m_t^2)] h_t^2 v_2 \right\} = 0,
\end{aligned} \quad (36)
$$

where

$$b(m^2) = m^2 (\ln \frac{m^2}{m_t^2} - 1).$$

From the minimization conditions one obtains:

$$
\begin{aligned}
v^2 &= \frac{4}{(g^2 + g^{'2})(\tan^2 \beta - 1)} \left\{ m_1^2 - m_2^2 \tan^2 \beta \right. \quad (37) \\
&- \left. \frac{3 h_t^2}{16\pi^2} \left[ [b(\tilde{m}_{t1}^2) + b(\tilde{m}_{t2}^2) - 2 b(m_t^2)] \tan^2 \beta + (A_t^2 m_0^2 \tan^2 \beta - \mu^2) \frac{b(\tilde{m}_{t1}^2) - b(\tilde{m}_{t2}^2)}{\tilde{m}_{t1}^2 - \tilde{m}_{t2}^2} \right] \right\} \\
2 m_3^2 &= (m_1^2 + m_2^2) \sin 2\beta + \frac{3 h_t^2 \sin 2\beta}{16\pi^2} \left\{ b(\tilde{m}_{t1}^2) + b(\tilde{m}_{t2}^2) - 2 b(m_t^2) \right. \\
&+ \left. (A_t m_0 + \mu \tan \beta)(A_t m_0 + \mu \cot \beta) \frac{b(\tilde{m}_{t1}^2) - b(\tilde{m}_{t2}^2)}{\tilde{m}_{t1}^2 - \tilde{m}_{t2}^2} \right\} \quad (38)
\end{aligned}
$$

From the above equations one can derive easily:

$$M_Z^2 = 2\frac{m_1^2 - m_2^2 \tan^2\beta - \Delta_Z^2}{\tan^2\beta - 1}, \tag{39}$$

$$\Delta_Z^2 = \frac{3g^2}{32\pi^2}\frac{m_t^2}{M_W^2\cos^2\beta}\left[b(\tilde{m}_{t1}^2) + b(\tilde{m}_{t2}^2) + 2m_t^2 + (A_t^2 m_0^2 - \mu^2\cot^2\beta)\frac{b(\tilde{m}_{t1}^2) - b(\tilde{m}_{t2}^2)}{\tilde{m}_{t1}^2 - \tilde{m}_{t2}^2}\right]$$

The Higgs masses corresponding to this one loop potential are [43, 44, 45, 46, 47, 23]:

$$m_A^2 = m_1^2 + m_2^2 + \Delta_A^2, \tag{40}$$

$$\Delta_A^2 = \frac{3g^2}{32\pi^2}\frac{m_t^2}{M_W^2\sin^2\beta}\left[b(\tilde{m}_{t1}^2) + b(\tilde{m}_{t2}^2) + 2m_t^2 + (A_t^2 m_0^2 + \mu^2)\frac{b(\tilde{m}_{t1}^2) - b(\tilde{m}_{t2}^2)}{\tilde{m}_{t1}^2 - \tilde{m}_{t2}^2}\right]$$

$$m_{H^\pm}^2 = m_A^2 + M_W^2 + \Delta_H^2, \tag{41}$$

$$\Delta_H^2 = -\frac{3g^2}{32\pi^2}\frac{m_t^4 \mu^2}{\sin^4\beta M_W^2}\frac{c(\tilde{m}_{t1}^2) - c(\tilde{m}_{t2}^2)}{\tilde{m}_{t1}^2 - \tilde{m}_{t2}^2}$$

$$m_{h,H}^2 = \frac{1}{2}\left[m_A^2 + M_Z^2 + \Delta_{11} + \Delta_{22}\right. \tag{42}$$

$$\left.\pm \sqrt{\begin{array}{ll}(m_A^2 + M_Z^2 + \Delta_{11} + \Delta_{22})^2 & -4m_A^2 M_Z^2\cos^2 2\beta - 4(\Delta_{11}\Delta_{22} - \Delta_{12}^2) \\ -4(\cos^2\beta M_Z^2 + \sin^2\beta M_A^2)\Delta_{22} & -4(\sin^2\beta M_Z^2 + \cos^2\beta M_A^2)\Delta_{11} \\ -4\sin 2\beta(M_Z^2 + M_A^2)\Delta_{12}\end{array}}\right]$$

$$\Delta_{11} = \frac{3g^2}{16\pi^2}\frac{m_t^4}{\sin^2\beta M_W^2}\left[\frac{\mu(A_t m_0 + \mu\cot\beta)}{\tilde{m}_{t1}^2 - \tilde{m}_{t2}^2}\right]^2 d(\tilde{m}_{t1}^2, \tilde{m}_{t2}^2),$$

$$\Delta_{22} = \frac{3g^2}{16\pi^2}\frac{m_t^4}{\sin^2\beta M_W^2}\left[\ln(\frac{\tilde{m}_{t1}^2 \tilde{m}_{t2}^2}{m_t^4}) + \frac{2A_t m_0(A_t m_0 + \mu\cot\beta)}{\tilde{m}_{t1}^2 - \tilde{m}_{t2}^2}\ln(\frac{\tilde{m}_{t1}^2}{\tilde{m}_{t2}^2})\right.$$

$$\left.+ \left[\frac{A_t m_0(A_t m_0 + \mu\cot\beta)}{\tilde{m}_{t1}^2 - \tilde{m}_{t2}^2}\right]^2 d(\tilde{m}_{t1}^2, \tilde{m}_{t2}^2)\right],$$

$$\Delta_{12} = \frac{3g^2}{16\pi^2}\frac{m_t^4}{\sin^2\beta M_W^2}\frac{\mu(A_t m_0 + \mu\cot\beta)}{\tilde{m}_{t1}^2 - \tilde{m}_{t2}^2}\left[\ln(\frac{\tilde{m}_{t1}^2}{\tilde{m}_{t2}^2}) + \frac{A_t m_0(A_t m_0 + \mu\cot\beta)}{\tilde{m}_{t1}^2 - \tilde{m}_{t2}^2}d(\tilde{m}_{t1}^2, \tilde{m}_{t2}^2)\right]$$

where

$$c(m^2) = \frac{m^2}{m^2 - \tilde{m}_q^2}\ln\frac{m^2}{\tilde{m}_q^2},$$

$$d(m_1^2, m_2^2) = 2 - \frac{m_1^2 + m_2^2}{m_1^2 - m_2^2}\ln\frac{m_1^2}{m_2^2}$$

and $\tilde{m}_q^2$ is the mass of a light squark.

## 5 Comparison of the MSSM with Data

In this chapter the various low energy GUT predictions are compared with data. The most restrictive constraints are the coupling constant unification and the requirement that the unification scale has to be above $10^{15}$ GeV from the proton lifetime limits, assuming decay via s-channel exchange of heavy gauge bosons. They exclude the SM [38, 5, 6] as well as many other models [5, 48, 49] with either a more complicated Higgs sector or models, in which one searches for the minimum number of new particles needed for unification. From the many models tried, only a few yielded unification at the required energies, but these models have particles introduced ad-hoc without the appealing properties of Supersymmetry. Therefore we will concentrate here on the supersymmetric models and ask if the predictions of the simplest, i.e. *minimal* models [50] are consistent with all constraints from data at low energy.

## 5.1 Unification of the Couplings

In the SM based on the group $SU(3) \times SU(2) \times U(1)$ the couplings are defined as:
$$\begin{array}{rcccl}
\alpha_1 & = & (5/3)g'^2/(4\pi) & = & 5\alpha/(3\cos^2\theta_W) \\
\alpha_2 & = & g^2/(4\pi) & = & \alpha/\sin^2\theta_W \\
\alpha_3 & = & g_s^2/(4\pi) &&
\end{array} \quad (43)$$

where $g'$, $g$ and $g_s$ are the $U(1)$, $SU(2)$ and $SU(3)$ coupling constants; the first two coupling constants are related to the fine structure constant by:
$$e = \sqrt{4\pi\alpha} = g\sin\theta_W = g'\cos\theta_W. \quad (44)$$

The factor of 5/3 in the definition of $\alpha_1$ has been included for the proper normalization at the unification point. The couplings, when defined as effective values including loop corrections in the gauge boson propagators, become energy dependent ("running"). A running coupling requires the specification of a renormalization prescription, for which one usually uses the modified minimal subtraction ($\overline{MS}$) scheme [37].

In this scheme the world averaged values of the couplings at the $Z^0$ energy are
$$\alpha^{-1}(M_Z) = 127.9 \pm 0.1 \quad (45)$$
$$\sin^2\theta_{\overline{MS}} = 0.2324 \pm 0.0005 \quad (46)$$
$$\alpha_3 = 0.123 \pm 0.006. \quad (47)$$

The value of $\alpha^{-1}$ is given in ref. [51] and the value of $\sin^2\theta_{\overline{MS}}$ has been been taken from a detailed analysis of all available data by Langacker and Polonsky [52], which agrees with the latest analysis of the LEP data [53]. The error includes the uncertainty from the top quark. We have not used the smaller error of 0.0003 for a given value of $m_t$, since the fit was only done within the SM, not the MSSM, so we prefer to use the more conservative error including the uncertainty from $m_t$.

The $\alpha_3$ value corresponds to the value at $M_Z$ as determined from quantities calculated in the "Next to Leading Log Approximation" [54]. These quantities are less sensitive to the renormalization scale, which is an indicator of the unknown higher order corrections; they are the dominant uncertainties in quantities relying on second order QCD calculations [55]. This $\alpha_s$ value is in excellent agreement with a preliminary value of $0.120 \pm 0.006$ from a fit to the $Z^0$ cross sections and asymmetries measured at LEP [53], for which the third order QCD corrections have been calculated too; the renormalization scale uncertainty is correspondingly small.

The top quark mass was simultaneously fitted to all electroweak data and found to be [53]:
$$M_{top} = 166^{+17}_{-19}{}^{+19}_{-22} \text{ GeV}, \quad (48)$$

where the first error is statistical and the second error corresponds to a variation of the Higgs mass between 60 and 1000 GeV. The central value corresponds to a Higgs mass of 300 GeV. This value is in good agreement with recent results quoted by the CDF Collaboration [32]:
$$M_{top} = 174^{+10}_{-10}{}^{+13}_{-12} \text{ GeV}, \quad (49)$$

where the first error is statistical and the second error systematic.

For SUSY models, the dimensional reduction $\overline{DR}$ scheme is a more appropriate renormalization scheme [56]. This scheme also has the advantage that all thresholds can be treated by simple step approximations. Thus unification occurs in the $\overline{DR}$ scheme if all three $\alpha_i^{-1}(\mu)$ meet exactly at one point. This crossing point then gives the mass of the heavy gauge bosons. The $\overline{MS}$ and $\overline{DR}$ couplings differ by a small offset

$$\frac{1}{\alpha_i^{\overline{DR}}} = \frac{1}{\alpha_i^{\overline{MS}}} - \frac{C_i}{12\pi} \qquad (50)$$

where the $C_i$ are the quadratic Casimir coefficients of the group ($C_i = N$ for SU($N$) and 0 for U(1) so $\alpha_1$ stays the same). Throughout the following, we use the $\overline{DR}$ scheme for the MSSM.

### 5.2  $M_Z$ from Electroweak Symmetry Breaking

In the MSSM at least two Higgs doublets have to be introduced. Radiative corrections from the heavy top and stop quarks can drive one of the Higgs masses negative, thus causing spontaneous symmetry breaking in the electroweak sector. In this case the Higgs potential does not have its minimum for all fields equal zero, but the minimum is obtained for non-zero vacuum expectation values of the fields. The scale, where symmetry breaking occurs depends on the starting values of the mass parameters at the GUT scale, the top mass and the evolution of the couplings and masses. This gives strong constraints between the known $Z^0$ mass and the SUSY mass parameters, as demonstrated e.g. in ref. [13].

After including the one-loop corrections to the potential [43, 44, 45, 46, 47, 23], the $M_Z$ mass becomes dependent on the top- and stop quark masses too (see eq. 39). Note that the corrections $\Delta_Z$ are zero if the top- and stop quark masses are identical, i.e. if supersymmetry would be exact. They grow with the difference $\tilde{m}_t^2 - m_t^2$, so these corrections become unnaturally large for large values of the stop masses, as will be discussed later.

### 5.3  $m_b$ from the $m_b/m_\tau$ Mass Ratio

Unification of the Yukawa couplings for a given generation at the GUT scale predicts relations for quark and lepton masses within a given family. Unfortunately, for the light quarks are the masses uncertain, but the ratio of b-quark and $\tau$-lepton masses can be correctly predicted by the radiative mass corrections [57, 12, 21, 22, 19, 20].

Assuming the simplest possible GUT model based on SU(5) gauge group, one has at the GUT scale: $m_b = m_\tau$. To calculate the experimentally observed mass ratio the RG equations for the running masses have to be used. By a physical mass we understand the value of the running mass at the energy scale equal to the mass itself. This definition of the mass is used throughout this paper.

From the RG equations for the Yukawa couplings one can easily obtain the RGE for the ratio

$$R_{b\tau} \equiv \frac{m_b}{m_\tau} = \sqrt{\frac{Y_b}{Y_\tau}},$$

see eq. 26 and table 4.

For the running mass of the b-quark we used [58]:

$$m_b = 4.25 \pm 0.3 \text{ GeV}. \qquad (51)$$

This mass depends on the choice of scale and the value of $\alpha_s(m_b)$. Consequently, we have assigned a rather conservative error of 0.3 GeV instead of the proposed value of 0.1 GeV [58]. Note that the running mass (in the $\overline{MS}$ scheme) is related to the physical (pole) mass $M_b^{pole}$ by [58]:

$$m_b = M_b^{pole}\left(1 - \frac{4}{3}\frac{\alpha_s}{\pi} - 12.4(\frac{\alpha_s}{\pi})^2\right) \approx 0.825\ M_b^{pole}, \tag{52}$$

so $m_b = 4.25$ corresponds to $M_b^{pole} \approx 5$ GeV. We ignore the running of $m_\tau$ below $m_b$ and use for the $\tau$ mass: $M_\tau = 1.7771 \pm 0.0005$ GeV [59].

### 5.4 Top Mass Constraints

The top mass can be expressed as:

$$m_t^2 = (4\pi)^2\ Y_t(t)\ v^2\ \sin^2(\beta), \tag{53}$$

where the running of the Yukawa coupling as function of $t = \log(\frac{M_{GUT}^2}{Q^2})$ in first order[4] is given by [12]:

$$Y_t(t) = \frac{Y_t(0)E(t)}{1 + 6Y_t(0)F(t)}, \tag{54}$$

where $E$ and $F$ are functions of the couplings only (see appendix). One observes that $Y_t(t)$ becomes independent of $Y_t(0)$ for large values of $Y_t(0)$, implying an upper limit on the top mass [14, 19, 20]. Requiring electroweak symmetry breaking implies a minimal value of the top Yukawa coupling, typically $Y_t(0) \geq \mathcal{O}(10^{-2})$. In this case the term $6Y_t(0)F(t)$ in the denominator of eq. (54) is much larger than one, since $F(t) \approx 290$ at the weak scale, where $t \approx 66$. In this case $Y_t(t) = E(t)/6F(t)$, so from eq. (53) it follows:

$$m_t^2 = \frac{(4\pi)^2\ E(t)}{6F(t)}\ v^2\ \sin^2(\beta) \approx (190\ \text{GeV})^2 \sin^2(\beta), \tag{55}$$

The physical (pole) mass is about 6% larger than the running mass [58]:

$$M_t^{pole} = m_t\left(1 + \frac{4}{3}\frac{\alpha_s}{\pi}\right) \approx (200\ \text{GeV})\sin\beta,. \tag{56}$$

The electroweak breaking conditions require $\pi/4 < \beta < \pi/2$ ; hence the equation above implies for the MSSM approximately:

$$145 < M_t^{pole} < 200\ \text{GeV}, \tag{57}$$

which is consistent with the experimental values given in eqns. (48) and (49).

For large top masses, the b-quark mass becomes a sensitive function of $m_t$ and of the starting values of the gauge couplings at $M_{GUT}$, as can be checked from the first order solution of the RGE for the ratio $R_{b\tau}$:

$$R_{b\tau}(m_b) = \left[\frac{\tilde{\alpha}_3(m_b)}{\tilde{\alpha}_3(m_Z)}\right]^{12/23}\left[\frac{\tilde{\alpha}_1(m_b)}{\tilde{\alpha}_1(m_Z)}\right]^{30/103} \tag{58}$$

---

[4]Throughout the analysis we have used the second order RG equations, for which no analytical solution exists, but this will not change the following arguments dramatically.

$$\times \quad \left[\frac{\tilde{\alpha}_3(m_Z)}{\tilde{\alpha}_3(m_t)}\right]^{12/23} \left[\frac{\tilde{\alpha}_1(m_Z)}{\tilde{\alpha}_1(m_t)}\right]^{15/53} \tag{59}$$

$$\times \quad \left[\frac{\tilde{\alpha}_3(m_t)}{\tilde{\alpha}_3(m_S)}\right]^{4/7} \left[\frac{\tilde{\alpha}_1(m_t)}{\tilde{\alpha}_1(m_S)}\right]^{10/41} [1 + 9/2 Y_t(t_S) F(t_t)]^{1/6} \tag{60}$$

$$\times \quad \left[\frac{\tilde{\alpha}_3(m_S)}{\tilde{\alpha}_3(m_G)}\right]^{8/9} \left[\frac{\tilde{\alpha}_1(m_S)}{\tilde{\alpha}_1(m_G)}\right]^{10/99} [1 + 6 Y_t(t_G) F(t_S)]^{-1/12}, \tag{61}$$

where the values of the Yukawa coupling at the GUT and SUSY scales, indicated by $m_G$ and $m_S$ or $t_G$ and $t_S$, respectively, are related to the Yukawa coupling at $m_t$:

$$Y_t(t_S) = \frac{Y_t(t_t)}{E(t_t) - 6 Y_t(t_t) F(t_t)}, \quad Y_t(t_G) = \frac{Y_t(t_S)}{E(t_S) - 9/2 Y_t(t_S) F(t_S)}.$$

The correlation between $m_t$ and $m_b$ originates from the $Y_t$ terms, which are large as will be shown later.

### 5.5 Experimental Lower Limits on SUSY Masses

SUSY particles have not been found so far and from the searches at LEP one knows that the lower limit on the charged leptons and charginos is about half the $Z^0$ mass (45 GeV) [31] and the Higgs mass has to be above 62 GeV [30]. The lower limit on the lightest neutralino is 18.4 GeV [31], while the sneutrinos have to be above 41 GeV [31]. These limits require minimal values for the SUSY mass parameters.

There exist also limits on squark and gluino masses from the hadron colliders [31], but these limits depend on the assumed decay modes. Furthermore, if one takes the limits given above into account, the constraints from the limits of all other particles are usually fulfilled, so they do not provide additional reductions of the parameter space in case of the *minimal* SUSY model.

### 5.6 Proton Lifetime Limits

GUT's predict proton decay and the present lower limits on the proton lifetime yield quite strong constraints on the GUT scale and the SUSY parameters. The direct decay $p \to e^+ \pi^0$ via s-channel exchange requires the GUT scale to be above $10^{15}$ GeV. This is not fulfilled in the SM, but always fulfilled in the MSSM. Therefore we do not consider this constraint. However, the decay via box diagrams with winos and Higgsinos predict much shorter lifetimes, especially in the preferred mode $p \to \bar{\nu} K^+$. From the present experimental lower limit of $10^{32}$ yr for this decay mode Arnowitt and Nath [28] deduce an upper limit on the parameter B:

$$B < (293 \pm 42) M_{H_3}/3 M_{\text{GUT}} \ GeV^{-1} \tag{62}$$

Here $M_{H_3}$ is the Higgs triplet mass, which is expected to be of the order of $M_{\text{GUT}}$. To obtain a conservative upper limit on $B$, we allow $M_{H_3}$ to become an order of magnitude heavier than $M_{\text{GUT}}$, so we require

$$B < 977 \pm 140 \ GeV^{-1}. \tag{63}$$

The uncertainties from the unknown heavy Higgs mass are large compared with the contributions from the first and third generation, which contribute through the mixing in the CKM matrix. Therefore we only consider the second order generation contribution, which can be written as [28]:

$$B = \frac{-2\alpha_2}{\alpha_3 \sin(2\beta)} \frac{m_{\tilde{g}}}{m_{\tilde{q}}^2} \, 10^6 \tag{64}$$

where $\alpha_2$ and $\alpha_3$ are the coupling constant of the $SU(2)$ and $SU(3)$ groups at the SUSY scale, respectively. One observes that the upper limit on $B$ favours small gluino masses $m_{\tilde{g}}$, large squark masses $m_{\tilde{q}}$, and small values of $\tan\beta$. To fulfill this constraint requires

$$\tan\beta < 10 \tag{65}$$

for the whole parameter space. Arnowitt and Nath note that requiring the gluino mass to be below 500 GeV implies the mass of the scalar particles ($m_0$) at the GUT scale to be above 600 GeV. We will not impose this requirement on the gluino mass. Furthermore, they require $M_{H_3} < 3\, M_{\text{GUT}}$, so they obtained tighter limits on $\tan\beta$, since we allow $M_{H_3} < 10\, M_{\text{GUT}}$.

## 5.7 Fit Strategy

As mentioned before, given the five parameters in the MSSM plus $\alpha_{\text{GUT}}$ and $M_{\text{GUT}}$, all other SUSY masses, the b-quark mass, and $M_Z$ can be calculated by performing the complete evolution of the couplings including all thresholds.

The proton lifetime limits prefer small values of $\tan\beta$, as discussed in the previous section, while all SUSY masses are expected to be below 1 TeV from the fine-tuning argument.

Therefore the following strategy was adopted: $m_0$ and $m_{1/2}$ were varied between 0 and 1000 GeV and $\tan\beta$ between 1 and 10. The trilinear coupling $A_t(0)$ at $M_{\text{GUT}}$ was kept mostly at zero, but the large radiative corrections to it were taken into account, so at lower energies it is unequal zero. Varying $A_t(0)$ between $+3m_0$ and $-3m_0$ did not change the results significantly, so the following results are quoted for $A_t(0) = 0$.

The remaining four parameters - $\alpha_{\text{GUT}}$, $M_{\text{GUT}}$, $\mu$, and $Y_t(0)$ - were fitted for each choice of $m_0, m_{1/2}$ and $\tan\beta$. Alternatively, fits were performed in which $m_{1/2}$ or $\tan\beta$ were left free too.

For unification in the $\overline{DR}$ scheme, all three couplings $\alpha_i^{-1}(\mu)$ must cross at a single unification point. Thus in these models one can fit the couplings at $M_Z$ by extrapolating from a single starting point at $M_{\text{GUT}}$ back to $M_Z$ for each of the $\alpha_i$'s and taking into account all light thresholds. The fitting program will then adjust the starting values of the four high energy parameters ($M_{\text{GUT}}$, $\alpha_{\text{GUT}}$, $\mu$ and $Y_t(0)$) until the five low energy values (three coupling constants, $M_Z$ and $m_b$) are "hit". The fit is repeated for all values of $m_0$ and $m_{1/2}$ between 100 and 1000 GeV and $\tan\beta$ between 1 and 10.

The light thresholds are taken into account by changing the coefficients of the RGE at the value $Q = m_i$, where the threshold masses $m_i$ are obtained from the analytical solutions of the corresponding RGE. These solutions depend on the integration range, which was chosen between $m_i$ and $M_{\text{GUT}}$. However, since one does not know $m_i$ at the beginning, an iterative procedure has to be used: one first uses $M_Z$ as a lower integration limit, calculates $m_i$, and uses this as lower limit in the next iteration. Of course, since the coupling constants are running, the latter have to be iterated too, so the values of $\alpha_i(m_i)$ have to be used for calculating the mass at the scale $m_i$ [13, 60]. Usually three to five iterations are enough to find a stable solution.

Following Ellis, Kelley and Nanopoulos [38] the possible effects from heavy thresholds are set to zero, since the proton lifetime limits forbid the Higgs triplet masses to be below $M_{\text{GUT}}$.

These heavy thresholds have been considered by other authors for different assumptions [61, 52, 62].

The most probable parameter values were obtained by minimizing the following $\chi^2$ function[5]:

$$\chi^2 = \sum_{i=1}^{3} \frac{(\alpha_i^{-1}(M_Z) - \alpha_{MSSM_i}^{-1}(M_Z))^2}{\sigma_i^2}$$
$$+ \frac{(M_Z - 91.18)^2}{\sigma_Z^2}$$
$$+ \frac{(m_b - 4.25)^2}{\sigma_b^2}$$
$$+ \frac{(B - 997)^2}{\sigma_B^2}(for\ B > 997)$$
$$+ \frac{(D(m1m2m3))^2}{\sigma_D^2}(for\ D > 0)$$
$$+ \frac{(\tilde{M} - \tilde{M}_{exp})^2}{\sigma_{\tilde{M}}^2}(for\ \tilde{M} > \tilde{M}_{exp}). \tag{66}$$

The first term is the contribution of the difference between the three calculated and measured gauge coupling constants at $M_Z$ and the following two terms are the contributions from the $M_Z$-mass and $m_b$-mass constraints. The last three terms impose constraints from the proton lifetime limits, from electroweak symmetry breaking, i.e. $D = V_H(v_1, v_2) - V_H(0,0) < 0$, and from experimental lower limits on the SUSY masses. The top mass, or equivalently, the top Yukawa coupling enters sensitively into the calculation of $m_b$ and $M_Z$. Instead of the top Yukawa coupling one could have taken the top mass as a parameter. However, if the couplings are evolved from $M_{\rm GUT}$ downwards, it is more convenient to run also the Yukawa coupling downward, since the RG equations of the gauge and Yukawa couplings form a set of coupled differential equations in second order, see eqs. (24) - (26). Once the Yukawa coupling is known at $M_{\rm GUT}$, the top mass can be calculated at any scale. In principle the experimental value of the top mass can be taken as a constraint too, as will be discussed in the next section.

The following errors were attributed: $\sigma_i$ are the experimental errors in the coupling constants, as given above, $\sigma_b$=0.3 GeV, $\sigma_B = 140$ GeV$^{-1}$, while $\sigma_D$ and $\sigma_{\tilde{M}}$ were set to 10 GeV. The values of the latter errors are not critical, since the corresponding terms in the numerator are zero in case of a good fit and even for the 90% C.L. limits these constraints could be fulfilled and the $\chi^2$ was determined by the other terms, for which one knows the errors.

In total one has to solve a system of 18 coupled differential equations: 5 second order ones (for the 3 gauge couplings, eq. (24), $Y_t$, eq. (25), and $Y_b/Y_\tau$, eq. (26)), and 13 first order ones (for the masses and parameters in the Higgs sector, see sect. 3). The second order ones are solved numerically[6] taking into account the thresholds of the light particles using the iteration procedure discussed above. Note that from the starting values of the parameters at $M_{\rm GUT}$ one can calculate all light thresholds from the simple first order equations before one starts the numerical integration of the five second order equations. Consequently, the program is fast in finding the optimum solution, even if before each iteration the light thresholds have to be recalculated.

---

[5]We use the MINUIT program from F. James and M. Roos, *MINUIT Function Minimization and Error Analysis*, CERN Program Library Long Writeup D506; Release 92.1, from March 1992. Our $\chi^2$ has discontinuities due to the experimental bounds on various quanitities, which become "active" only for specific regions of the parameter space. Consequently the derivatives are not everywhere defined. The option SIMPLEX, which does not rely on derivatives, can be used to find the monotonous region and the option MIGRAD to optimize inside this region.

[6]The program DDEQMR from the CERN library was used for the solution of these coupled second order differential equations.

## 5.8 Results

We first consider fits without proton lifetime constraints, since they are more questionable due to the uncertainty of the Higgs masses near the GUT scale. The upper part of fig. 2 shows the evolution of the coupling constants in the MSSM for two cases: one for the minimum value of the $\chi^2$ function given in eq. 66 (solid lines) and one corresponding to the 90% C.L. upper limit of the thresholds of the light SUSY particles (dashed lines). The position of the light thresholds is shown in the bottom part as jumps in the first order $\beta$ coefficients, which are increased according to the entries in table 1 as soon as a new threshold is passed. Also the second order coefficients are changed correspondingly (see table 2), but their effect on the evolution is not visible in the top figure in contrast to the first order effects, which change the slope of the lines considerably in the top figure. One observes that the changes in the coupling constants occur in a rather narrow energy regime, so qualitatively this picture is very similar to the case, in which all sparticles were assumed to be degenerate at an effective SUSY mass scale $M_{\rm SUSY}$ [5]. Since the running of the couplings depends only logarithmically on the sparticle masses, the 90% C.L. upper limits are as large as several TeV, as shown by the dashed lines in fig. 2 and more quantitatively in table 5. With the fitted SUSY parameters given at the top of the table, the corresponding masses of the SUSY particles can be calculated. Their values are given in the lower part of the table. The upper and lower limits in table 5 will be discussed below.

The mass of the lightest Higgs particle, called $h$ in table 5, is a rather strong function of $m_t$, as shown in fig. 3 for various choices of $\tan\beta$, $m_0$ and $m_{1/2}$. All other parameters were optimized for these inputs and after the fit the values of the Higgs and top mass were calculated and plotted. One observes that the mass of the lightest Higgs particle varies between 60 and 150 GeV and the top mass between 134 and 190 GeV. Furthermore, it is evident that $\tan\beta$ almost uniquely determines the value of $m_t$ (through eq. 53), since even if $m_{1/2}$ and $m_0$ are varied between 100 and 1000 GeV, one finds practically the same $m_t$ for a given $\tan\beta$. The value of $m_t$ varies between 134 and 190 GeV, if $\tan\beta$ is varied between 1.2 and 10. This range is in excellent agreement with the estimates given in eq. 57, if one takes into account that $M_t^{pole} \approx 1.06 m_t$ (see eq. 56).

As mentioned before, varying $A_t(0)$ between $+3m_0$ and $-3m_0$ does not influence the results very much, so its value at the unification scale was kept at 0, but its non-zero value at lower energies due to the large radiative corrections was taken into account. The fits are shown for positive values of the Higgs mixing paramter $\mu$, but similar values are obtained for negative values of $\mu$ with an equally good $\chi^2$ value for the fit.

The parameters $m_0$, $m_{1/2}$ and $\mu$ are correlated, as shown in fig. 4, where the value of $\mu$ is shown for all combinations of $m_0$ and $m_{1/2}$ between 100 and 1000 GeV. One observes that $\mu$ increases with increasing $m_0$ and $m_{1/2}$.

The $\chi^2$ slowly increases with increasing values of $\mu$ and $m_{1/2}$, as shown in fig. 5. The steep walls originate from the experimental lower limits on the SUSY masses and the requirement of radiative symmetry breaking. In the minimum the $\chi^2$ value is zero, but one notices a long valley, where the $\chi^2$ is only slowly increasing. Consequently, the upper limits on the sparticle masses, which grow with increasing values of $\mu$ and $m_{1/2}$, become several TeV, as shown in table 5. The 90% C.L. upper limits were obtained by requiring an increase in $\chi^2$ of 1.64.

The correlation between $m_{1/2}$ and $\mu$ originates mainly from the electroweak symmetry breaking condition, but also from the fact that the thresholds in the running of the gauge couplings all have to occur at a similar scale. For example, from fig. 2 it is obvious that the dashed lines for $1/\alpha_1$ and $1/\alpha_2$ from the 90% C.L. will not meet with the solid line of $1/\alpha_3$, simply because the thresholds are too different; the thresholds in $1/\alpha_3$ are mainly determined by $m_{1/2}$, while the thresholds for the upper two lines include the winos and higgsinos too, so one obtains automatically a positive correlation between $\mu$ and $m_{1/2}$.

The upper limits on the SUSY masses are a sensitive function of the central value of $\alpha_s$: decreasing the central value of $\alpha_s$ by two standard deviations (i.e. 0.012) can increase the thresholds of sparticles several TeV. Acceptable fits are only obtained for input $\alpha_s$ values between 0.108 and 0.132, if the error is kept at 0.006. Outside this range all requirements cannot be met simultaneously any more, so the MSSM predicts $\alpha_s$ in this range.

As discussed previously, sparticle masses in the TeV range spoil the cancellation of the quadratic divergences. This can be seen explicitly in the corrections to $M_Z$ (eq. 39): $\Delta_Z$ is exactly zero if the masses of stop– and top quarks are identical, but the corrections grow quickly if the degeneracy is removed, as shown in fig. 6. For the SUSY masses at the minimum value of $\chi^2$ the corrections to $M_Z$ are small. If one requires that only solutions are allowed for which the corrections to $M_Z$ are not large compared with $M_Z$ itself, one has to limit the mass of the heaviest stop quark to about one TeV. The corresponding 90% C.L. upper limits of the individual sparticles masses are given in the right hand column of table 5. The correction to $M_Z$ is 6 times $M_Z$ in this case. The limits are obtained by scanning $m_0$ and $m_{1/2}$ till the $\chi^2$ value increases by 1.64, while optimizing the values of $\tan\beta$, $\mu$, , $\alpha_{\text{GUT}}$, $Y_t(0)$ and $M_{\text{GUT}}$. The lower limits on the SUSY parameters are shown in the left column of table 5. The lowest values of $m_0$ and $m_{1/2}$ are required to have simultaneously a sneutrino mass above 42 GeV and a wino mass above 45 GeV. If the proton lifetime limit is included, either $m_0$ or $m_{1/2}$ have to be above a certain limit as will be discussed below. Since the squarks and gauginos are much more sensitive to $m_{1/2}$ than $m_0$, one obtains the lower limits by increasing $m_0$. The minimum value for $m_0$ is about 400 GeV in this case. But in both cases the $\chi^2$ increase for the lower limits is due to the b-mass, which is predicted to be 4.6 GeV from the parameters determining the lower limits, so it gives a contributions to the $\chi^2$ function, which requires $m_b$=4.25 GeV (see eq. 51).

The b-quark mass is a strong function of both, $\tan\beta$ and $m_t$, as shown in fig. 7; this dependence originates from the $W - t$ loop to the bottom quark. The horizontal band corresponds to the running mass of the b-quark (eq. 51): $m_b = 4.25 \pm 0.3$ GeV. As shown in fig. 3, the top mass is strongly correlated with $m_t$; consequently the correlation between $m_b$ and $m_t$ depends on $\tan\beta$, as shown by the two curves in fig. 7. Since also $M_Z$ is a strong function of the same parameters, the requirement of gauge and Yukawa coupling unification together with electroweak symmetry breaking strongly constrains the SUSY particle spectrum. A typical fit with a $\chi^2$ equal zero is given in the central column of table 5, but is should be noted that the values in the other columns provide acceptable fits too at the 90% C.L. .

In fig. 3 only the unification and electroweak symmetry breaking constraints were taken into account, not the ones from the proton lifetime limits. The proton lifetime constraint (eq. 64) requires $m_{1/2}$ and $m_0$ to be above a certain minimum value, as can be seen from fig. 8: the preferred region, i.e. the lowest $\chi^2$ is obtained for values away from the origin. The increase in the corner is completely due to the constraint from the proton lifetime. This plot was made for $\tan\beta = 2$. For larger values the region excluded by proton decay quickly increases; for $\tan\beta = 10$ practically the whole region is excluded, as shown in fig. 9. The recent estimates from the CDF Collaboration for the top mass (eq. 49) are shown as the shaded area in fig. 3. If one includes this constraint in the $\chi^2$ definition by adding a term $(M_t^{pole} - 174)^2/17^2$ one finds:

$$1.2 < \tan\beta < 5.5 \text{ at the 90\% C.L.} \tag{67}$$

for the whole parameter space.

# 6  Summary

The MSSM model has many predictions, which can be compared with experiment, even in the energy range where the predicted SUSY particles are out of reach. Among these predictions:

- $M_Z$.
- $m_b/m_\tau$.
- Proton decay.
- Limits on $m_t$.

It is surprising, that the *minimal* supersymmetric model can fulfil all experimental constraints for these predictions. As far as we know, supersymmetric models are the only ones, which are consistent with all these observations simultaneously. Other models can yield unification too [48], but they do not exhibit the elegant symmetry properties of supersymmetry, they offer no explanation for dark matter and no explanation for the electroweak symmetry breaking. Furthermore the quadratic divergencies do not cancel.

As mentioned in the introduction, this is the first analysis using a $\chi^2$ definition to determine the probability of each point in the SUSY parameter space, which allows us to determine 90% C.L. limits and the most probable values ( see table 5 ). The fit to all data simultaneously yields at the 90% C.L. the following parameter ranges (if extreme finetuning is to be avoided, i.e. $\tilde{m}_{t2} < 1$ TeV and excluding the proton lifetime constraint):

$$
\begin{aligned}
65 &< m_0 &< 1000 \text{ GeV} \\
37 &< m_{1/2} &< 475 \text{ GeV} \\
117 &< |\mu| &< 1100 \text{ GeV} \\
1.2 &< \tan\beta &< 5.5 \\
145 &< M_t^{pole} &< 200 \text{ GeV (from fig. 3.)} \\
0.108 &< \alpha_s &< 0.132
\end{aligned}
$$

The upper limit on $m_0$ originates from the fine-tuning constraint, the upper limit on $\tan\beta$ from the top mass estimates. If the proton lifetime limit is considered too, the lower limit on $m_0$ would be 250 GeV (see fig. 9). The fact that $\tan\beta$ is so much smaller than the ratio of the top– and b-quark masses implies that the Yukawa coupling of the b-quark is negligibly small, so one does not have to consider its contributions in the RG equations.

Good fits are only obtained for $\alpha_s$ between 0.108 and 0.132, if the error on $\alpha_s$ is taken to be 0.006. The bottom mass constraint together with the given couplings require the top mass to be between 134 and 190 GeV, in perfect agreement with the experimental values.

From the allowed ranges for the parameters one finds the corresponding constraints on the SUSY masses (see table 5 for details):

$$\begin{aligned}
18 &< \chi_1^0(\tilde{\gamma}) < 202 \text{ GeV} \\
39 &< \chi_2^0(\tilde{Z}), \chi_1^\pm(\tilde{W}) < 386 \text{ GeV} \\
109 &< \tilde{g} < 1104 \text{ GeV} \\
115 &< \tilde{q} < 1070 \text{ GeV} \\
140 &< \tilde{t}_1 < 725 \text{ GeV} \\
218 &< \tilde{t}_2 < 1000 \text{ GeV} \\
82 &< \tilde{e}_L < 521 \text{ GeV} \\
67 &< \tilde{e}_R < 440 \text{ GeV} \\
41 &< \tilde{\nu}_L < 516 \text{ GeV} \\
109 &< \chi_3^0(\tilde{H}_1) < 799 \text{ GeV} \\
120 &< \chi_4^0(\tilde{H}_2) < 812 \text{ GeV} \\
129 &< \chi_2^\pm(\tilde{H}^\pm) < 831 \text{ GeV} \\
121 &< H^\pm < 1034 \text{ GeV} \\
118 &< H < 1033 \text{ GeV} \\
92 &< A < 1031 \text{ GeV} \\
60 &< h < 142 \text{ GeV} \quad \text{(from fig. 3.)}
\end{aligned}$$

The lower limits will all increase as soon as the LEP limits on sneutrinos, winos and the lightest Higgs increase. The lightest Higgs particle is certainly within reach of experiments at present or future accelerators [63, 64]. Its observation in the predicted mass range of 60 to 140 GeV would be a strong case in support of this minimal version of a supersymmetric grand unified theory.

It should be noted that for the typical fit in table 5 the mass of the lightest stop is 20 GeV lower the top mass, mainly due to the low value of $\tan\beta$ and relatively large value of $\mu$, see eq. (21). In this case the stop cannot decay into the top, but can decay as follows:

$$\tilde{t}_1 \to \tilde{\chi}_1^\pm + b \to \tilde{\chi}_1^0 + W + b \to \tilde{\chi}_1^0 + lepton + \nu + b,$$

which is experimentally very similar to the normal top decay signature [65]. Additional stop production could be an explanation for the excess of events seen by the CDF Collaboration: they observe an effective cross section for top pair production of $13.9^{+6.1}_{-4.8}$ pb, while the calculated $t\bar{t}$ cross section is only $5.8^{+0.8}_{-0.4}$ pb [32]. In case of a larger top mass $\tan\beta$ increases (see fig. 3), thus leading to a weaker mixing, in which case the stop mass would be above the top mass.

**Acknowledgement.**
We thank G. Altarelli, M. Drees and T. Sjöstrand for discussions.

**APPENDIX A**

We present here the notation used above:

$$E(t) = (1 + \beta_3 t)^{16/(3b_3)}(1 + \beta_2 t)^{3/b_2}(1 + \beta_1 t)^{13/(15b_1)}$$

$$\begin{aligned}
F(t) &= \int_0^t E(t')dt' \\
f_i(t) &= \frac{1}{\beta_i}\left(1 - \frac{1}{(1+\beta_i t)^2}\right) \\
h_i(t) &= \frac{t}{(1+\beta_i t)}, \quad \beta_i = b_i\,\tilde{\alpha}_{\text{GUT}} \\
e(t) &= \frac{3}{2}\left[\frac{G_1(t)+Y_0 G_2(t)}{D(t)} + \frac{(H_2(t)+6Y_0 H_4(t))^2}{3D^2(t)} + H_8(t)\right] \\
f(t) &= -\frac{6Y_0 H_3(t)}{D^2(t)} \\
h(t) &= \frac{1}{2}(\frac{3}{D(t)} - 1) \\
k(t) &= \frac{3Y_0 F(t)}{D^2(t)} \\
q(t) &= \frac{1}{(1+6Y_0 F(t))^{1/4}}(1+\beta_2 t)^{3/(2b_2)}(1+\beta_1 t)^{3/(10b_1)} \\
r(t) &= \left(\frac{3Y_0 H_3(t)}{D(t)} - H_7(t)\right)q(t) \\
s(t) &= \frac{3Y_0 F(t)}{D(t)}q(t) \\
D(t) &= 1 + 6Y_0 F(t) \\
H_2(t) &= \tilde{\alpha}_{\text{GUT}}(\frac{16}{3}h_3(t) + 3h_2(t) + \frac{13}{15}h_1(t)) \\
H_3(t) &= tE(t) - F(t) \\
H_4(t) &= F(t)H_2(t) - H_3(t) \\
H_5(t) &= \tilde{\alpha}_{\text{GUT}}(-\frac{16}{3}f_3(t) + 6f_2(t) - \frac{22}{15}f_1(t)) \\
H_6(t) &= \int_0^t H_2^2(t')E(t')dt' \\
H_7(t) &= \tilde{\alpha}_{\text{GUT}}(3h_2(t) + \frac{3}{5}h_1(t)) \\
H_8(t) &= \tilde{\alpha}_{\text{GUT}}(-\frac{8}{3}f_3(t) + f_2(t) - \frac{1}{3}f_1(t)) \\
G_1(t) &= F_2(t) - \frac{1}{3}H_2^2(t) \\
G_2(t) &= 6F_3(t) - F_4(t) - 4H_2(t)H_4(t) + 2F(t)H_2^2(t) - 2H_6(t) \\
F_2(t) &= \tilde{\alpha}_{\text{GUT}}(\frac{8}{3}f_3(t) + \frac{8}{15}f_1(t)) \\
F_3(t) &= F(t)F_2(t) - \int_0^t E(t')F_2(t')dt' \\
F_4(t) &= \int_0^t E(t')H_5(t')dt'
\end{aligned}$$

| Symbol | Lower limits | | **Typical fit** | 90% C.L. Upper limits | |
|---|---|---|---|---|---|
| Constraints | GEY | GEY+P | **GEY+(PF)** | GEY+ (P) | GEY+(P)+F |
| Fitted SUSY parameters | | | | | |
| $m_0$ | 65 | 400 | **400** | 400 | 400 |
| $m_{1/2}$ | 37 | 80 | **111** | 1600 | 475 |
| $\mu$ | -117 | 330 | **870** | 1842 | 1101 |
| $\tan\beta$ | 3.0 | 3.0 | **1.56** | 8.5 | 2.9 |
| $Y_t(0)$ | 0.0158 | 0.0035 | **0.0150** | 0.0023 | 0.0084 |
| $M_t^{pole}$ | – | – | **175** | 178 | 189 |
| $m_t$ | – | – | **165** | 168 | 178 |
| $1/\alpha_{\rm GUT}$ | 23.8 | 24.3 | **24.5** | 25.9 | 25.2 |
| $M_{\rm GUT}$ | $2.3\ 10^{16}$ | $2.0\ 10^{16}$ | **$2.0\ 10^{16}$** | $0.8\ 10^{16}$ | $1.3\ 10^{16}$ |
| SUSY masses in [GeV] | | | | | |
| $\chi_1^0(\tilde\gamma)$ | 18 | 25 | **41** | 720 | 202 |
| $\chi_2^0(\tilde Z)$ | 39 | 52 | **80** | 1346 | 386 |
| $\chi_1^\pm(\tilde W)$ | 46 | 48 | **79** | 1347 | 386 |
| $\tilde g$ | 109 | 217 | **293** | 3377 | 1105 |
| $\tilde e_L$ | 82 | 406 | **409** | 1160 | 521 |
| $\tilde e_R$ | 67 | 401 | **402** | 729 | 440 |
| $\tilde\nu_L$ | 41 | 400 | **406** | 1157 | 516 |
| $\tilde q_L$ | 120 | 443 | **477** | 3030 | 1071 |
| $\tilde q_R$ | 115 | 440 | **471** | 2872 | 1030 |
| $\tilde b_L$ | 112 | 352 | **369** | 2610 | 903 |
| $\tilde b_R$ | 119 | 440 | **471** | 2862 | 1027 |
| $\tilde t_1$ | – | – | **144** | 2333 | 725 |
| $\tilde t_2$ | – | – | **467** | 2817 | 1008 |
| $\chi_3^0(\tilde H_1)$ | 109 | 292 | **540** | 1771 | 799 |
| $\chi_4^0(\tilde H_2)$ | 120 | 313 | **556** | 1780 | 812 |
| $\chi_2^\pm(\tilde H^\pm)$ | 129 | 315 | **566** | 1816 | 831 |
| $h$ | – | – | **87** | 146 | 127 |
| $H$ | 118 | 523 | **812** | 2218 | 1033 |
| $A$ | 92 | 521 | **810** | 2217 | 1031 |
| $H^\pm$ | 121 | 527 | **813** | 2219 | 1034 |

Table 5: Values of SUSY masses and parameters for various constraints: G=gauge coupling unification; E=electroweak symmetry breaking; Y=Yukawa coupling unification; P=Proton lifetime constraint; F=finetuning constraint. Constraints in brackets indicate that they are fulfilled but not required. The minimum values of the lightest Higgs mass, the stop mass and the top mass can't be reached for the parameters minimizing the squarks and slepton masses. One needs smaller values of $\tan\beta$ in that case.

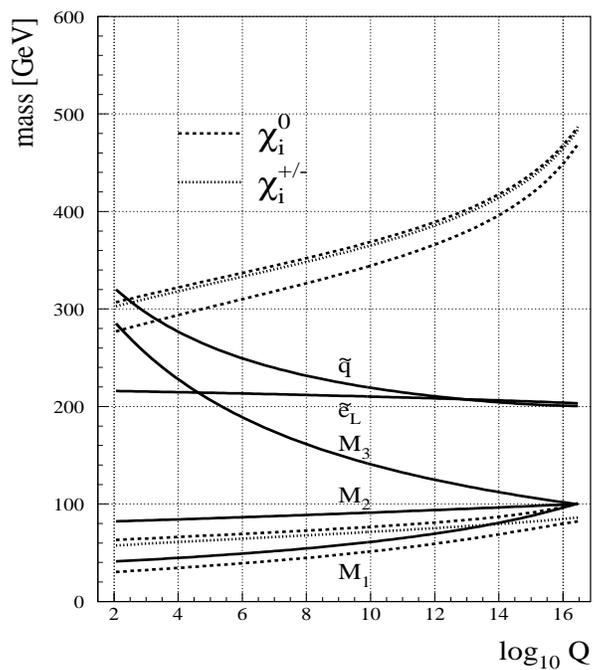

Figure 1: Typical running of the squark ($\tilde{q}$), slepton ($\tilde{e}_L$), and gaugino ($M_1$, $M_2$, $M_3$) masses (solid lines). The dashed lines indicate the running of the four neutralinos and two charginos.

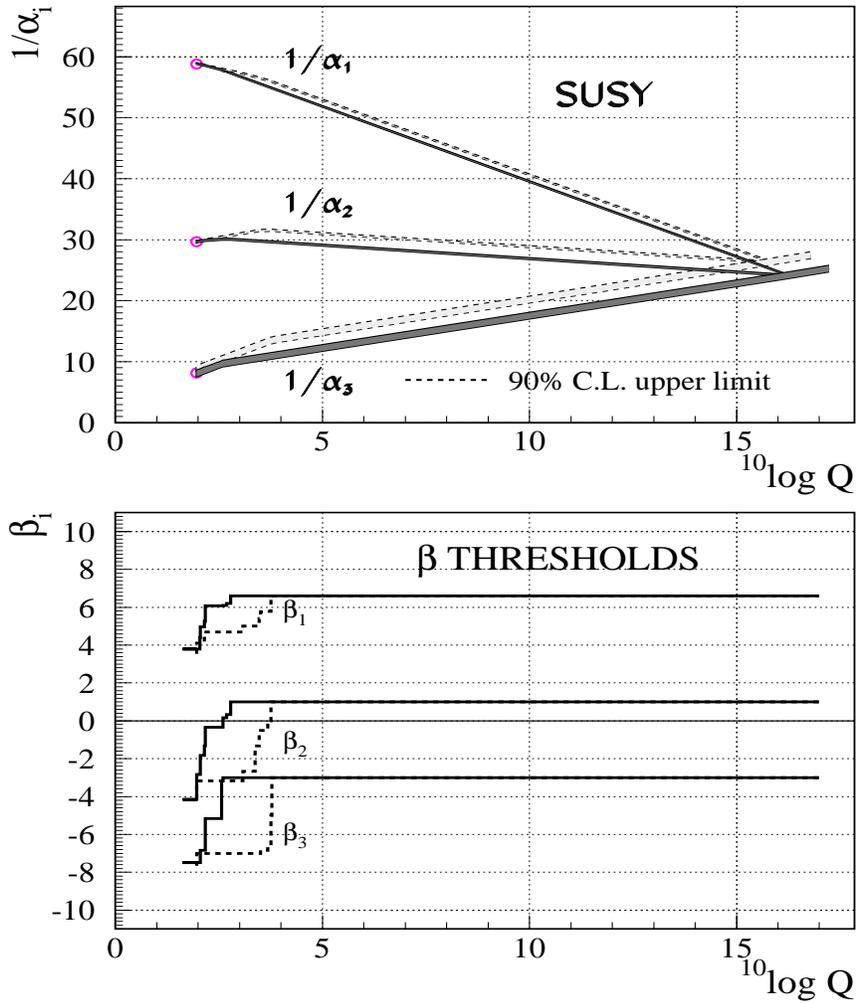

Figure 2: Evolution of the inverse of the three couplings in the MSSM. The line above $M_{\mathrm{GUT}}$ follows the prediction from the supersymmetric SU(5) model. The SUSY thresholds have been indicated in the lower part of the curve: they are treated as step functions in the first order $\beta$ coefficients in the renormalization group equations, which correspond to a change in slope in the evolution of the couplings in the top figure. The dashed lines correspond to the 90% C.L. upper limit for the SUSY thresholds.

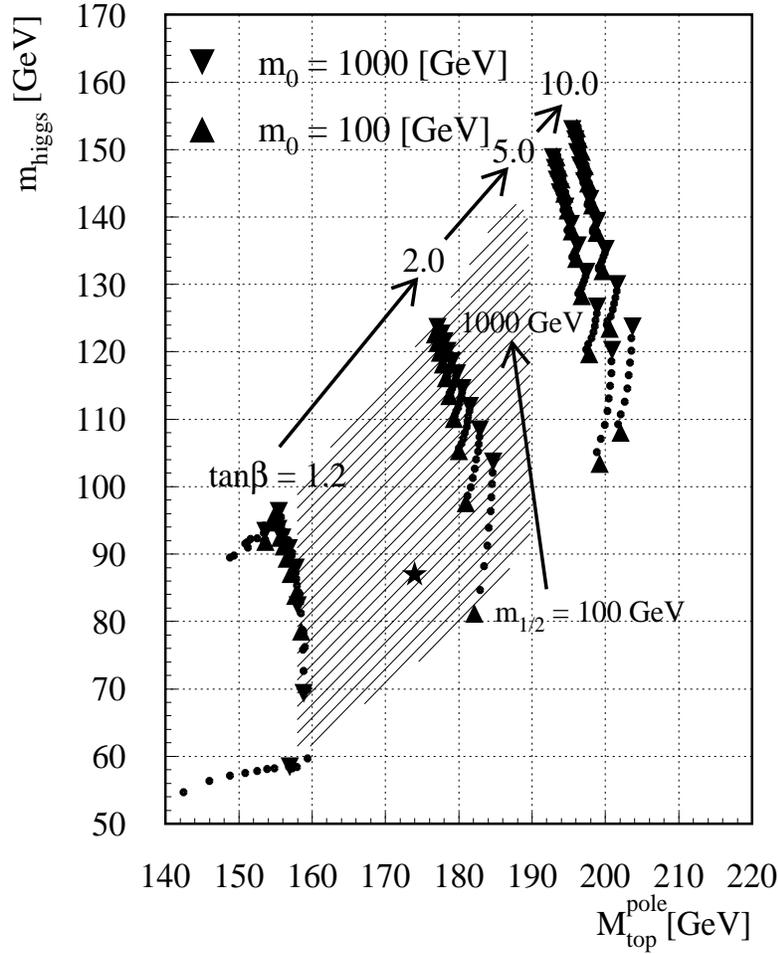

Figure 3: The mass of the lightest Higgs particle as function of the top quark mass for values of $\tan\beta$ between 1.2 and 10 and values of $m_0$ and $m_{1/2}$ between 100 and 1000 GeV. The parameters of $\mu$, $M_{\text{GUT}}$, $\alpha_{\text{GUT}}$ and $Y_t(0)$ are optimized for each choice of these parameters; the corresponding values of the top and lightest Higgs mass are shown as symbols. For small values of $m_{1/2}$ the Higgs mass increases with $m_0$, as shown for a "string" of points, each representing a step of 100 GeV in $m_0$ for a given value of $m_{1/2}$, which is increasing in steps of 100 GeV, starting with the low values for the lowest strings. At high values of $m_{1/2}$ the value of $m_0$ becomes irrelevant and the "string" shrinks to a point. Note the strong positive correlation between $m_{higgs}$ and all other parameters: the highest value of the Higgs mass corresponds to the maximum values of the input parameters, i.e. $\tan\beta = 10$, $m_0 = m_{1/2} = 1000$ GeV; this value does not correspond to the minimum $\chi^2$. More likely values are: $m_{higgs} \approx 87$ GeV for $m_{1/2} = 100$ GeV, $m_0 = 400$ GeV, $\mu = 822$ GeV and $\tan\beta = 1.6$, as indicated by the star. The hatched area corresponds to the top mass range measured by [32].

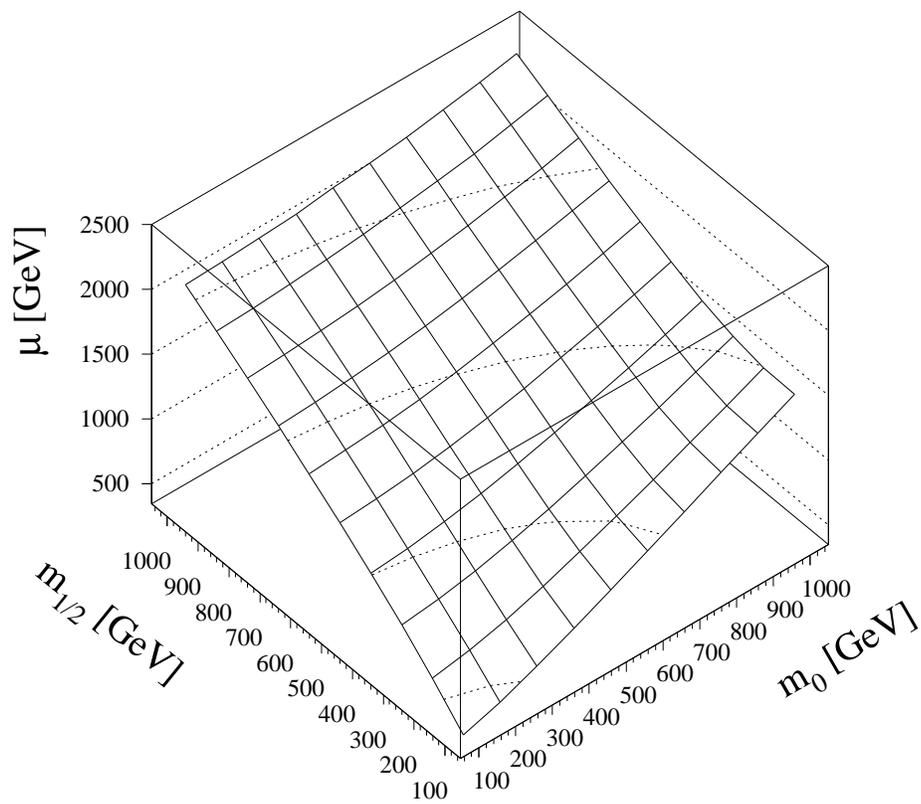

Figure 4: The fitted MSSM parameter $\mu$ as function of $m_0$ and $m_{1/2}$ for $\tan\beta = 2$.

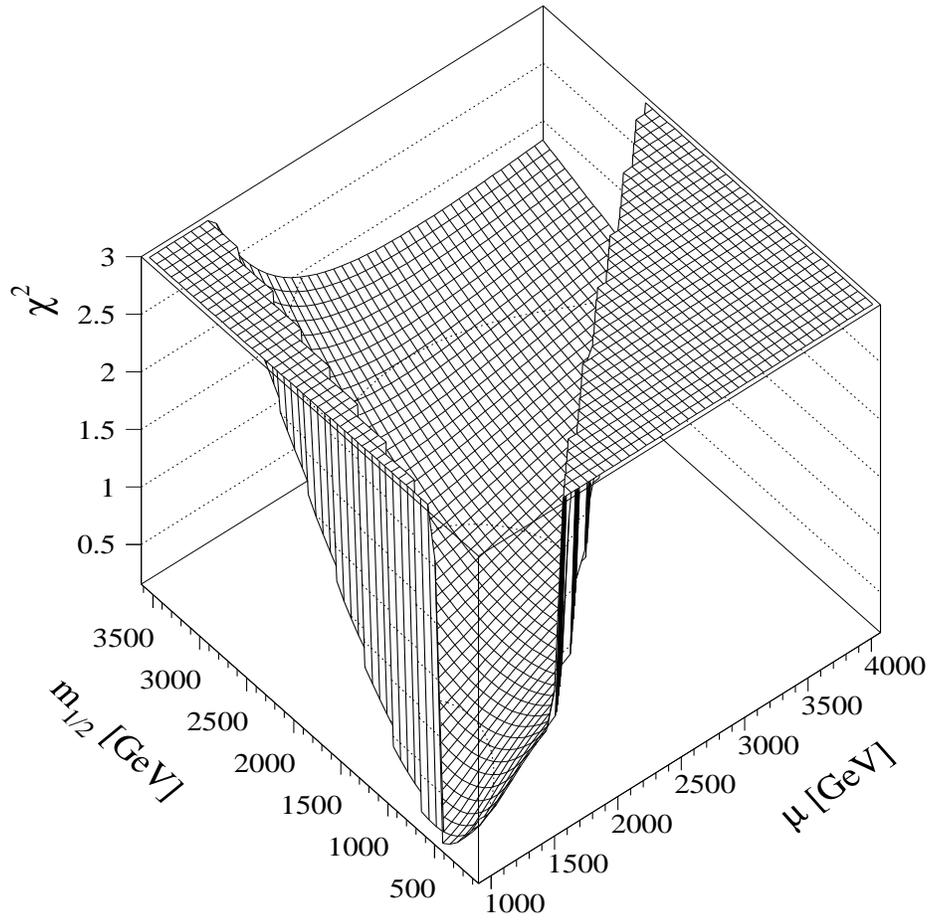

Figure 5: The correlation between $m_{1/2}$ and $\mu$ for $m_0=500$ GeV.

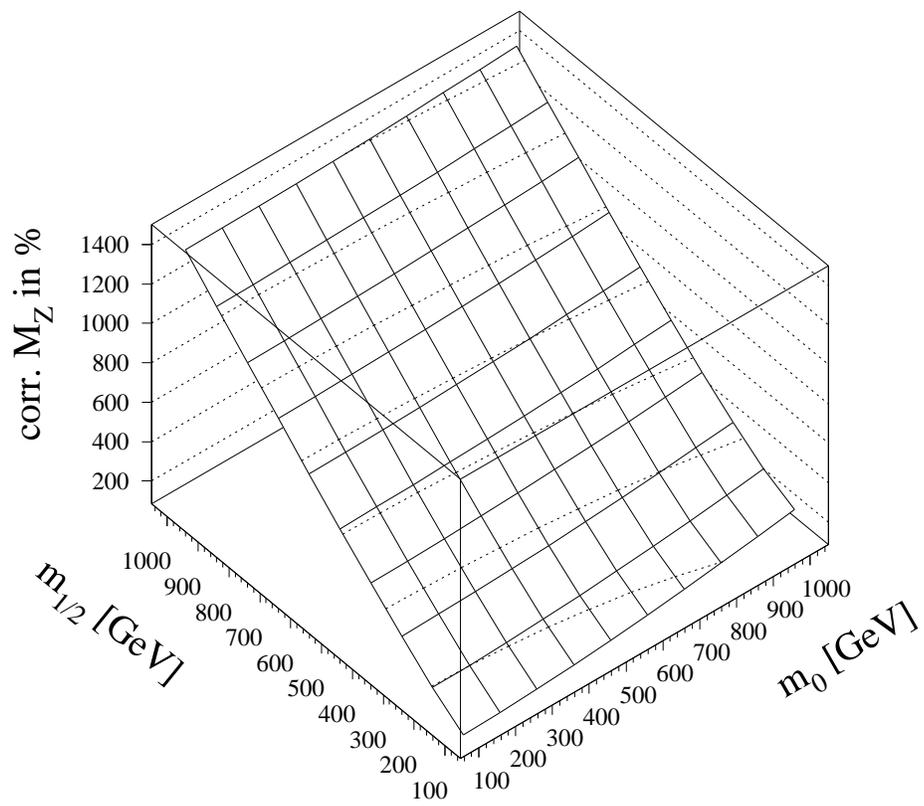

Figure 6: The one-loop correction factor to $M_Z$ as function of $m_0$ and $m_{1/2}$.

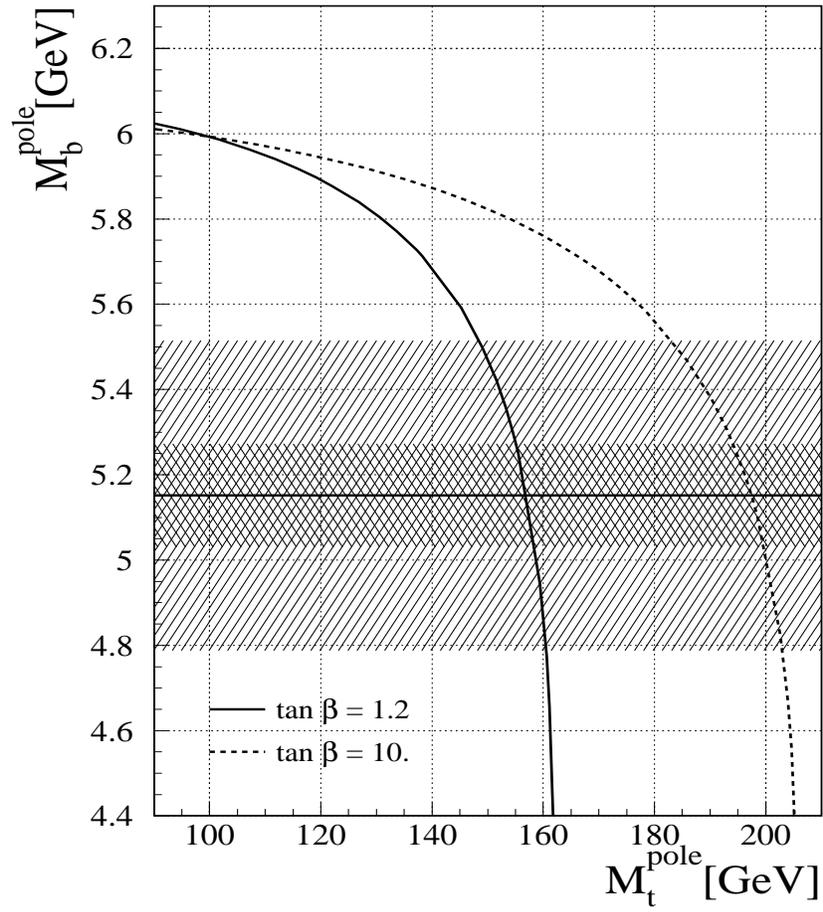

Figure 7: The correlation between the physical pole masses $M_b^{pole}$ and $M_t^{pole}$ for $m_0$=400 GeV and two values of $\tan\beta$. The hatched area indicates the experimental value for $M_b^{pole}$.

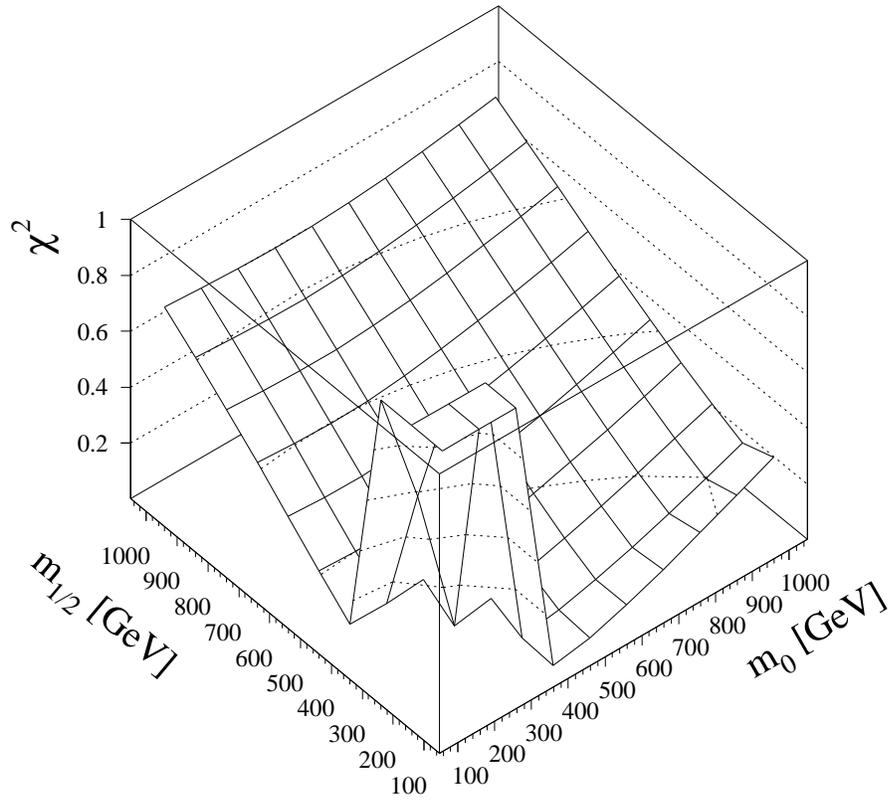

Figure 8: The $\chi^2$ of the fit as function of $m_0$ and $m_{1/2}$ for $\tan\beta = 2$. The sharp increase in $\chi^2$ in the corner is caused by the lower limit on the proton lifetime.

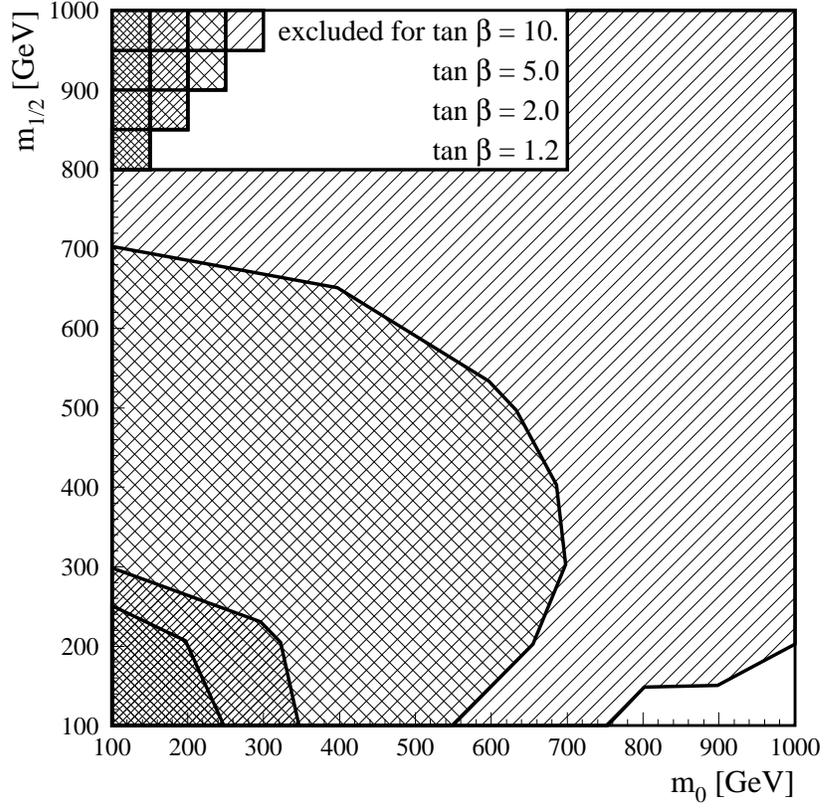

Figure 9: The 90% C.L. contours from proton lifetime limits for different values of $\tan\beta$. For $\tan\beta = 1.2$ only the small corner on the lower left part is excluded (small $\tilde{q}$ masses), whereas for $\tan\beta = 10$ nearly the whole region is excluded. Only the lower right corner (small $m_{1/2}$ and large $m_0$) is then still allowed. Consequently the proton decay requires $\tan\beta < 10$ for practically the whole parameter space.